\documentclass{jpp}
\usepackage{graphicx}
\usepackage{natbib}
\usepackage{color}
\usepackage{amsmath}

\ifCUPmtlplainloaded \else
  \checkfont{eurm10}
  \iffontfound
    \IfFileExists{upmath.sty}
      {\typeout{^^JFound AMS Euler Roman fonts on the system,
                   using the 'upmath' package.^^J}%
       \usepackage{upmath}}
      {\typeout{^^JFound AMS Euler Roman fonts on the system, but you
                   dont seem to have the}%
       \typeout{'upmath' package installed. JPP.cls can take advantage
                 of these fonts, if you use 'upmath' package.^^J}%
      }
  \else
  \fi
\fi

\ifCUPmtlplainloaded \else
  \checkfont{msam10}
  \iffontfound
    \IfFileExists{amssymb.sty}
      {\typeout{^^JFound AMS Symbol fonts on the system, using the
                'amssymb' package.^^J}%
       \usepackage{amssymb}%
       \let\le=\leqslant  
       \let\ge=\geqslant  
      }{}
  \fi
\fi

\ifCUPmtlplainloaded \else
  \IfFileExists{amsbsy.sty}
    {\typeout{^^JFound the 'amsbsy' package on the system, using it.^^J}%
     \usepackage{amsbsy}}
    {\providecommand\boldsymbol[1]{\mbox{\boldmath $##1$}}}
\fi

\newcommand\J{$J$}            % Airy function

\newcommand{\Jshape}{\textsf{J}}
\newcommand{\Sshape}{\textsf{S}}

\newsavebox{\astrutbox}
\sbox{\astrutbox}{\rule[-5pt]{0pt}{20pt}}

\newcommand{\BE}{\begin{equation}}
\newcommand{\EE}{\end{equation}}
\newcommand{\BA}{\begin{eqnarray}}
\newcommand{\EA}{\end{eqnarray}}
 \newcommand{\fig}[1]{Fig.~\ref{fig_#1}}

 \newcommand{\sect}[1]{Sect.~\ref{sect_#1}}
 
 \newcommand{\eq}[1]{Eq.~(\ref{eq_#1})}

\newcommand{\ie}{\textit{i.e.}}

\newcommand\etc{etc.\ }
\newcommand\eg{e.g.}

\title[3D magnetic reconnection and its application to solar flares]{3D magnetic reconnection and its application to solar flares}

\author[M. Janvier]%
{M\ls I\ls H\ls O\ns J\ls A\ls N\ls V\ls I\ls E\ls R$^1$%
  \thanks{Email address for correspondence: miho.janvier@ias.u-psud.fr},\ns
}

\affiliation{$^1$Institut d'Astrophysique Spatiale, CNRS, Univ. Paris-Sud, Universit\'e Paris-Saclay, B\^{a}t. 121, 91405 Orsay cedex, France\\[\affilskip]
}

\pubyear{2010}
\volume{650}
\pagerange{119--126}
\date{?; revised ?; accepted ?. - To be entered by editorial office}
\begin{document}

\maketitle

\begin{abstract}
Solar flares are powerful radiations occuring in the Sun's atmosphere. They are powered by magnetic reconnection, a phemonenon that can convert magnetic energy into other forms of energy such as heat and kinetic energy, and it is believed to be ubiquitous in the universe.
With the ever increasing spatial and temporal resolutions of solar observations, as well as numerical simulations benefiting from increasing computer power, we can now probe into the nature and the characteristics of magnetic reconnection in 3D to better understand its consequences during eruptive flares in our star's atmosphere. 
We review in the following the efforts made on different fronts to approach the problem of magnetic reconnection. In particular, we will see how understanding the magnetic topology in 3D helps locating the most probable regions for reconnection to occur, how the current layer evolves in 3D and how reconnection leads to the formation of flux ropes, plasmoids and flaring loops.
\end{abstract}

\begin{PACS}
Authors should not enter PACS codes directly on the manuscript, as these must be chosen during the online submission process and will then be added during the typesetting process (see http://www.aip.org/pacs/ for the full list of PACS codes)
\end{PACS}

%________________________________________________________________________
%%%%%%%%%%%%%%%%%%%%%%%%%%%%%%%%%%%%%%%%%%%%%%%%%%%%%%%%%%%%%%%%%%%%%%%%%%%%%%%%%%%%%
\section{Introduction} %%%%%%%%%%%%%%%%%%%%%%%%%%%%%
\label{sect_Introduction}
  
%   {\S\bf Flares - generalities} \\

Solar flares are the most energetic events taking place in our solar system. They are detected as a sudden increase in the X-ray light emission as they take place in our Sun's atmosphere. Their energy ranges from $10^{24}$ erg (the detection lower limit) to $10^{32}$ erg \citep{Schrijver2012}, or in Joules $10^{17}$ J to $10^{25}$ J. They can then be classified or ranked with the intensity of the light curve peak in soft X-rays, as recorded by the Geostationary Operational Environmental Satellites (GOESs) near Earth. Their emissions are recorded in a wide range of the electromagnetic spectrum, from gamma rays and X-rays to radio wavelengths.
With a varied range of instruments aboard spacecraft and on the ground, our Sun can be studied with an incredible amount of details. 

While {less powerful flares (A,B and C-class flares) are most} often confined flares (their influence on the corona remains localised), {other flares of} higher intensities (so-called M- and X-class flares) {can be} responsible for the release of large clouds of solar plasma - called coronal mass ejections (CMEs) \citep[\eg][]{Yashiro2006}- in the interplanetary medium. These are also detected in the interplanetary medium (we then referred to them as interplanetary CMEs). They perturb the ambient solar wind and have characteristic signatures in terms of magnetic field, proton temperature, composition, ionisation ratio different than the ambient medium.
Relativistic solar energetic particles, accelerated from the Sun during flares, as well as CMEs, are important drivers of space weather, since they impact the magnetic environments of planets \citep[see][]{Gosling1991,Prange2004}. Then, understanding the underlying mechanisms of these eruptive flares is of primary importance to better assess the likelihood and the evolution of CMEs in the interplanetary medium. Understanding the flaring mechanism also questions the habitability of exoplanets orbiting around other stars.\\

%  {\S\bf Magnetic reconnection for flare modelling }\\
  
Since most of the energy available in the Sun's corona is in magnetic form (low-$\beta$ plasma), and as large solar flares mainly occur in the locations of sunspots, it is natural to suppose that solar flares are powered by magnetic energy. Then, there is a need for a mechanism that can convert this magnetic energy into other forms of energy such as heat, particle energies, etc. Such a work started with Cowling and Sweet \citep[among others, see \eg][]{Cowling1945,Sweet1950} who investigated the separation of the motion of the magnetic field and the plasma. Later on, Dungey \citep{Dungey1953} continued this work, proposing the existence of a surface where a strong Ohmic electric field exists due to the decrease of the conductivity of the medium, thus allowing changes in the ``identity" of field lines. This change of identity can be illustrated by magnetic field lines changing their connection with one another, hence the term magnetic reconnection. 

Several years later, Sweet \citep{Sweet1956} and Parker \citep{Parker1957b} proposed the formation of a current layer near magnetic nulls, where the magnetic field can be dissipated (see also \sect{topology}). Such a mechanism was put forward to explain the acceleration of solar particles to relativistic speeds \citep{Parker1957a,Sweet1958c}. Indeed, although it is impossible to directly probe the Sun's atmosphere to measure the magnetic field, consequences of magnetic reconnection were first observed as populations of energetic particles (solar cosmic rays) and radio bursts \citep[as summarized in \eg][]{Wild1963}. Theoretical models continue to be confronted with observations. For example, intensely emitting magnetic arcades loops and loop tops seen during flares \citep[\eg][]{Masuda1994}, coronal mass ejections \citep{Webb2012} and plasmoids \citep[{which can be defined as a circular plasma structure seen on a projection on the plane of the sky, \eg}][]{Takasao2012} are indicative of magnetic structures that are similar to what is found in simulations. Moreover, inflows and flare ribbons sweeping the Sun's surface \citep[see references in][]{McKenzie2011} are also believed to be {indirect} observable consequences of reconnecting magnetic fields. \\

%  {\S\bf Reconnection modelling at small scales}\\

Then, since Sweet's and Parker's seminal work in laying the basis for a magnetic reconnection model, scenarios explaining how this phenomenon actually takes place have flourished. First, we can distinguish between models that are focussed on the process of reconnection itself. These models, with consideration of topology and/or energetics, can have a more or less complete description of the Ohm's law \citep[including the Hall term and/or other kinetic effects at ion or electron scales, see \eg][]{Birn2001a}, and provide analyses of the evolution of the dissipation layer, which can become unstable (\eg\ the tearing instability) or into which the magnetic field can be forced to flow in and reconnect. Over the years, many papers have been dedicated to the understanding of these fundamental processes at various scales, and such studies are extremely valuable to better assess the energetics of observed flares. Dedicated plasma experiments \citep[such as MRX,][]{Yamada1997} and recent space missions \citep[\eg\ CLUSTER and MMS,][]{Sergeev2008,Fuselier2016} also shed some light on the evolution of the magnetic field, plasma and nonthermal particle populations during reconnection events. As will be seen in the following, magnetic reconnection remains a highly investigated domain due to its complexity. Because of the multiple approaches and the different highlights (kinetic effects, topology, energy, particles, ...), a review covering all these aspects is out the scope of the present work. We propose instead to the reader the reviews of \citet{Yamada2007,Zweibel2009} and \citet{Yamada2010} for more information on aspects of reconnection at the diffusion region scale, in the magnetosphere and in laboratory experiments.\\

%  {\S\bf Application to flares: the CSHKP model }\\

With the first observations of flares \citep[see the well-documented early report by][]{Severny1964}, several models started to emerge from the middle of the 20th century \citep[see][and references therein]{Sweet1969}. Of particular interest here are global models which aid in understanding the global evolution of the magnetic field and generic consequences that are observable. These models were initially 2D \citep[for example, the so-called CSHKP model, see][]{Carmichael1964, Sturrock1966, Hirayama1974,Kopp1976}. They evolved along the years to include more and more physical aspects of flares (heating of loops, energetic particles, \etc), as observations were made with better temporal and spatial resolutions. 
In this 2D model, the flare is powered by magnetic reconnection at a vertical current sheet forming in the corona below the ejected large-scale magnetic structure (or upward-propagating plasmoid{, where here the plasmoid refers to a region where the magnetic field takes a circular shape, in a 2D projection}) that forms the CME. Particles accelerated from, or waves generated at, the reconnection region, travel along field lines and impact/heat the chromosphere, the lowest and densest layer of the Sun's atmosphere. Localised heating is then seen as flare ribbons, that are believed to form at the footpoints of newly reconnected field lines, or coronal loops, which themselves are heated to extremely high temperatures (they can be seen in soft X-rays) and filled with dense plasma upon ``evaporation'' from the chromosphere. \\

%  {\S\bf Transition in 3D}\\

However, since the nature of flares is intrinsically 3D, those models are not able to completely describe certain features of flares. For example, a self-consistent model describing the evolution of the ejected magnetic structure, from its formation to its further expansion as well as its flux increase, is still lacking. Note that the ejected structure is often modelled as a flux rope, which can be described as a bundle of magnetic field lines twisting around each other. Furthermore, the evolution of flare loops displays a gradual transition from strongly sheared flare loop arcades to nearly potential configuration \citep[\eg][]{Asai2003,Su2006,Su2007,Warren2011}, while the morphology of flare ribbons is also peculiar \citep[with a  \J-shape, see Fig.6 in][]{Chandra2009}. Most importantly, the nature of reconnection in 3D is also not addressed in the CSHKP model and similar models.
Over the past few years, 3D numerical simulations have helped our understanding of the evolution of an unstable flux rope/magnetic configuration \citep[\eg][]{Aulanier2010,Aulanier2012,Kusano2012}, the evolution of the current layer and reconnection in 3D \citep{Janvier2013,Kliem2013} as well as its consequences for interpreting solar flares \citep[\eg][]{Dudik2014}.\\

%  {\S\bf Road map} \\
In the following, we review certain aspects of magnetic reconnection applied to the understanding of eruptive flares. We are aware that related reviews already exist \citep[\eg][]{Pontin2012}, although generally these are focused on the analytical/simulation side or observational side. With the recent advances in understanding reconnection in 3D and the growing evidence of observational data from solar flares confirming evidence laid by theoretical works, we thought beneficial to review their parallel evolution in the following. We also put them in a general context of reconnection studied in different fields of plasma studies. As a first step, we will see in \sect{topology} how the magnetic topology allows one to investigate the regions were magnetic reconnection is most likely to occur, such as null points, separatrices, or more generally, quasi-separatrix layers, which generalise the concepts of separatrices in 3D.
We will also investigate the different diffusion mechanisms that can take place in the current layer in \sect{currents} (\eg\ the onset of instabilities), as well as the current layer formation, evolution and observation in eruptive flares. The consequences of reconnection on the magnetic field (slipping of field lines, formation of flare loops, flux rope and plasmoids) are investigated in \sect{consequences}.
 Finally, in \sect{Conclusion}, we summarize our results, discuss the limitations of present studies of reconnection applied to solar flares, and conclude.

%%%%%%%%%%%%%%%%%%%%%%%%%%%%%%%%%%%%%%%%%%%%%%%%%%%%%%%%%%%%%%%%%%%%%%%%%%%%%%%%%%%%%
\section{Magnetic topology of solar flares} %%%%%%%%%%%%%%%%%%%%%%%%%%%%%
\label{sect_topology}

Solar flares are believed to be powered by a physical phenomenon called magnetic reconnection. When reconnection occurs, magnetic field lines rearrange their connectivity so as to create new magnetic structures. Then, two questions arise: what is this region where reconnection takes place? and where does it form? 

\subsection{Null points, separatrices and separators}  %%%%%%%%%%%%%%%%%%%%%%%%%%%%%
\label{sect_NP}

%   {\S\bf Beginnings} \\

\begin{figure}  %________________________ FIG ______________________________________    
\centering
\includegraphics[width=1\textwidth,clip]{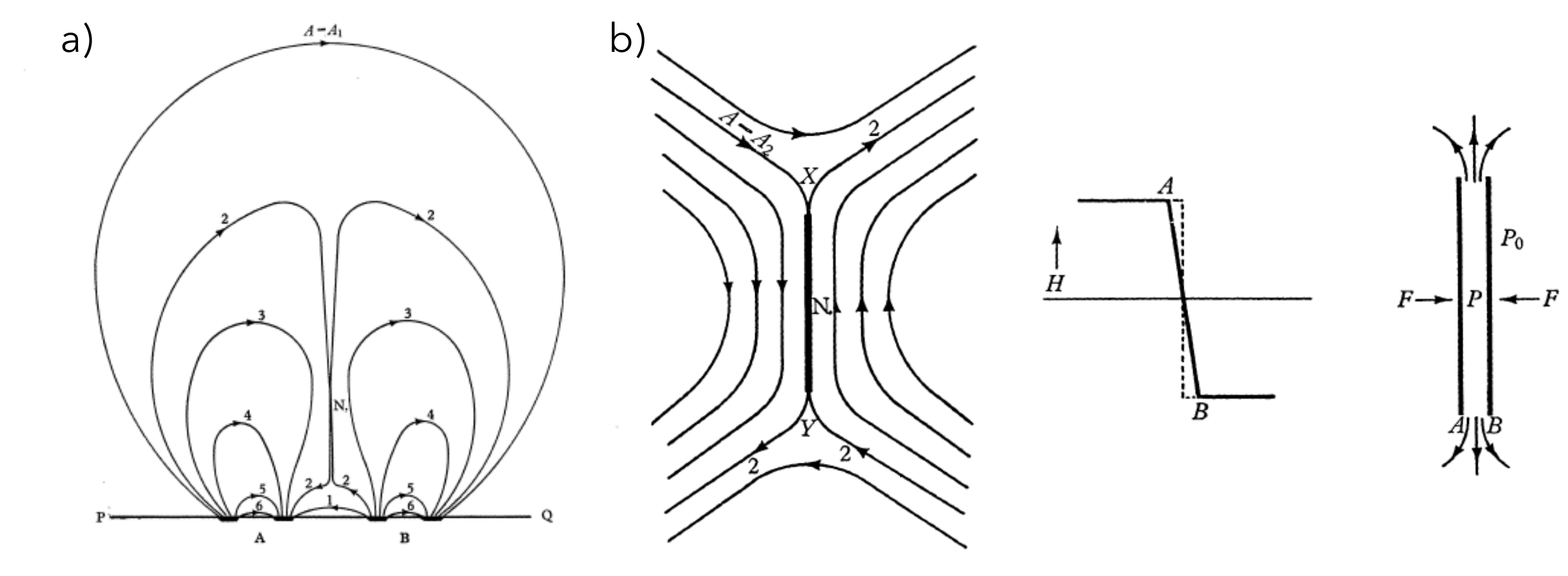}
\caption{Early analytical models of reconnection regions: (a) Sweet's mechanism at play when two bipolar sunspots are brought close to each other, forming a current layer surrounding the null point N1 (b) Sweet and Parker ÇÊcollision layerÊÈ (current sheet), with a description of the magnetic field (here called H), and the hydrodynamic model (most left panel). Adapted from \citealt{Sweet1958a}.
}
\label{fig_1}
\end{figure}

In the 1950s, the idea emerged that some plasma locations can be related with the magnetic field losing its identity: in other words, regions where the magnetic field decorrelates from the plasma. Sweet and Parker \citep[as summarised in][]{Parker1957b} provided a mechanism for the spontaneous formation of a diffusion layer, even in highly conducting media. Magnetic fields with opposite directions, when pressed together by external forces, form a surface where the electric current density is large (\fig{1}b). Subsequently, the gradient in the field density becomes so important that the diffusion terms, generally of negligible importance away from this singular layer, are responsible for the rapid diffusion of the magnetic field (see \sect{dissipationprocess} for a more detailed description of the different dissipation processes).

 %  {\S\bf When magnetic null points first appear} \\

Sweet's and Parker's proposal of tangential discontinuities was soon after extended by Sweet \citep{Sweet1958a} to the presence of magnetic nulls in a volume (following the concept of \citealt{Giovanelli1947}), where the magnetic field vanishes. Null points were seen as physically interesting in the concept of flares, as they were thought to offer the best locations for particle acceleration (as these locations would be associated with parallel electric field).
%, which counteracts the inhibiting action of magnetic field in particle acceleration). 
Sweet proposed that null points result from flux tubes protuding from sunspots. The distorted magnetic field in their vicinity then allows magnetic reconnection to occur. This work, along with that of \citet{Sweet1958d}, was important in laying the basis of the study of the topology of the magnetic field and the exchange of flux between different magnetic connectivity domains traced by the lines of magnetic force. {As null points are locations where the magnetic field line mapping is discontinuous, ideal kinematic solutions lead to a region where the magnetic field must be rapidly dissipated: this concept of current layers around null points was introduced by \citep{Syrovatskii1971}.}

%   {\S\bf The stability of null points - formation of current layers} \\

Magnetic null points are associated with current layers, \ie\ regions where the electric current density increases, either from a spontaneous formation due to some instabilities (\sect{currents}) or from a dynamic formation as a result of the magnetic field evolution. 
Analyses of the flow pattern near the null point reveal some geometrical properties of the current layer such as its length in quasistatic field evolutions \citep[see \eg][]{Syrovatskii1966, Priest1975}. Then, the consequences of magnetic reconnection in those singular regions were investigated analytically in more details, although the first attempts were mainly treating the problem as a boundary layer associated with some singular structure in a nearly ideal plasma \citep[\eg][]{Sonnerup1970, Yeh1970}. 

%   {\S\bf Topological definitions } \\

\begin{figure}  %________________________ FIG ______________________________________    
\centering
\includegraphics[width=1\textwidth,clip]{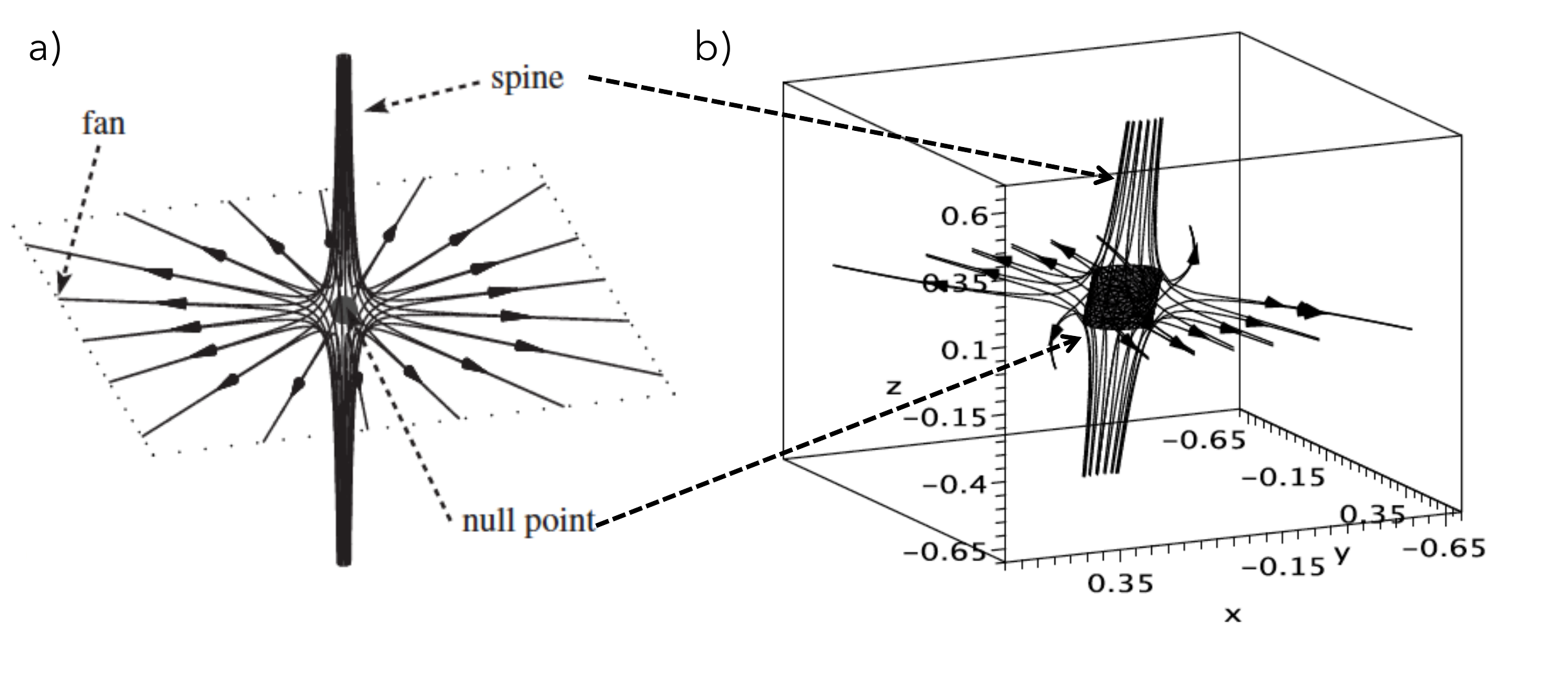}
\caption{Different topological definitions in the presence of a null point:  specific field lines that pass through the null form the spine, then they spread in the fan plane (defined by the two eigenvectors with sign-like eigenvalues). (a) represents a symmetric case, associated with two equal eigenvalues in the fan plane \citep{Pontin2012}, while (b) shows an asymmetric null point \citep[adapted from][]{AlHachami2010}. In the present plots, the field lines are selected to pass nearby the spine and the fan.
}
\label{fig_2}
\end{figure}

Studies of kinematic reconnection at null points were soon extended from 2D to 3D \citep[\eg][]{Yeh1976,Baum1979,Lau1990}.  This led to a definition of various topological objects, such as separators, which connect pairs of null points of opposite signs. The sign is defined with the highest number of positive or negative eigenvalues: since $\nabla \cdot \vec{B}=0$, two eigenvalues are positive and one is negative, or the reverse. In the presence of a separator, the current layer tends to form along it. Then, separatrices are lines (or surfaces in 3D) separating different domains of connectivity. Note that separatrices are also locations of current sheet formation when the magnetic field is sheared \citep[\eg][]{Vekstein1991}. 
Properties of nulls can also be found in \citet{Lau1990}, who described the three eigenvalues and eigenvectors of the magnetic field near the null point. Then, there is a specific field line that passes through the null, called the spine, which is associated to the eigenvalue with the single sign. All the other field lines connecting the null form a surface called the fan, associated with the two eigenvalues of same sign \citep[see \fig{2} and][]{Priest1996}. 
Overall, topological connectivity was regarded as essential as it contains information regarding the location and the structure where localised electric currents form during an evolution of the magnetic configuration.

%   {\S\bf Reconnection surrounding a null point} \\

{Separatrices and separators, similar to null points, are regions of discontinuous field line mapping. Its implication in the formation of discontinuities in the plasma, for example in the electric field, can be investigated by ideal kinematic solutions. Then,} over the years, different regimes of reconnection have been studied, {from a kinematic approach to a dynamic approach with the help of numerical simulations. D}epending on whether the process takes place at null points \citep[see][and references therein for more explanations on fan, spine reconnection regimes]{Parnell1996, Priest1996, Pontin2004, Pontin2005, Pontin2013}, or at separators, \ie\ with currents concentrated along a separator field line \citep{Longcope1996, Heerikhuisen2004, Pontin2006, Parnell2010, Stevenson2015}. The papers of \citet{Priest2009} and \cite{Priest2016} provide an overview of the different reconnection regimes happening around the null point as well as a link to different works on this topic. They also provided an extension to the classical definition of reconnection at spine/fan locations by defining three regimes: torsional spine, torsional fan and spine-fan reconnection. The two former occur when rotational motions (of either the spine or the fan) lead to the concentration of current along the spine or the fan, while the latter is when the current is concentrated along both fan and spine due to the shearing of a null point.
Although the early works were kinematic studies of reconnection around null points, over the years computer studies have made possible dynamic studies of reconnection and evolution of the current sheet near a null point \citep[\eg][]{Pontin2007,Galsgaard2011},  comparison with observations \citep[\eg][]{Mandrini1991,Masson2009}, as well as investigation beyond the classical MHD frame \citep[\eg\ in collisionless plasmas,][]{Tsiklauri2007}.\\

%   {\S\bf Null points in laboratory experiments} \\
   
Dynamics surrounding a null point can also be investigated in laboratory experiments. In \citet{Syrovatskii1972,Baum1973, Baum1976}, the authors showed that the plasma resistivity indeed becomes anomalously large, and the current becomes concentrated near null points. Recent laboratory experiments such as in \citet{Stenzel2002} teach us that the small scale physics involving ion and electron dynamics must also be carefully considered in the surrounding of null points. Finally, in \citet{Yamada2015}, the authors review recent understandings from laboratory experiments in the mechanisms responsible for the heating and the acceleration of ions as well as in the energy deposition near X-points, for the electrons (for which the energy mostly comes from the electric field component perpendicular to the magnetic field).

%    {\S\bf Null points in magnetic charge models} \\

The occurence of solar flares predominantly in regions of sunspots has driven the analysis of the magnetic field geometry and the current systems predominantly in those regions. 
%Sunspots were proposed as resulting from the emergence of flux tubes, and with the increasing resolutions of magnetograms, were priviledged locations to study the magnetic geometry of flaring regions. 
With the method proposed by \citet{Schmidt1964}, introducing a point charge to describe a magnetic field, and with the advent of the computer, kinematics of the plasma associated with different topology could then be studied numerically and be compared with observations. In \citet{Baum1980}, the authors studied numerically in details a system made of separatrices and null points, which allowed for comparisons with analytical works, and opened the door to proper studies of reconnection phenomena in the context of solar physics.

This model, referred to as a Magnetic Charge Topology (MCT) model, in the sense that it imposes distinct unipolar regions, was refined over the years \citep[\eg][]{Henoux1987}. It explained some flare features such as the two chromospheric ribbon-shape structures appearing during certain flares \citep{Gorbachev1988}. Then, more complex models with several charge sources can be introduced \citep{Mandrini1991, Demoulin1993, Barnes2005}, so as to compare for example the locations of the flare kernels (\ie\ brightened locations) and the topology \citep[see][and references therein]{Longcope2005}.

\begin{figure}  %________________________ FIG ______________________________________    
\centering
\includegraphics[width=1\textwidth,clip]{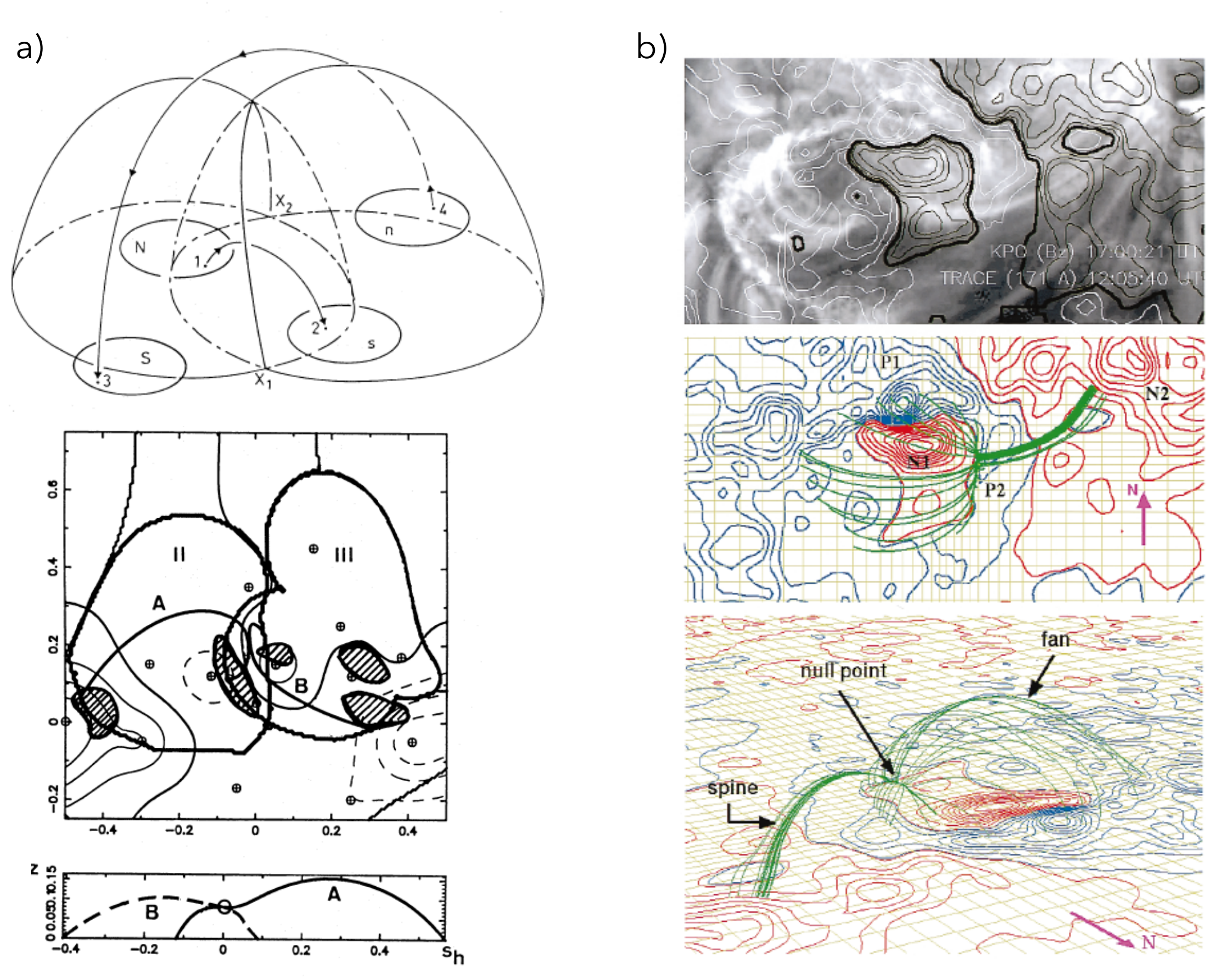}
\caption{(a) Simplified schema of separatrice surfaces in the presence of two bipoles 1-2 and 3-4 (top) and comparison of their photospheric traces with the locations of H$\alpha$ brightenings seen during a flare, adapted from \citet{Mandrini1991} (b) Magnetic field configuration of a flaring region associated with a spine and fan structure as obtained from a magnetic field extrapolation. They can directly be compared with emissions seen in extreme UV of coronal loops, adapted from \citet{Aulanier2000}.
}
\label{fig_3}
\end{figure}

Other methods, such as flux tubes \citep{Sakurai1977}, pointwise mapping or submerged poles/dipoles models \citep[\eg][]{Demoulin1994} have also been investigated, with similar purposes to compare the complex topology features with observed features (see \fig{3}). Chromospheric kernels and flare ribbons were found to appear on a part of the photopheric footprints of separatrices\citep[\eg][]{Mandrini1993, Henoux1993, Demoulin1994}. They provided, over the years and with an increasing number of observations and refined techniques, convincing support for the hypothesis that magnetic energy is indeed being converted during solar flares via magnetic reconnection. \\

%Coming back to applications to the solar dynamical atmosphere, it is possible to investigate the topology of the magnetic field in the Sun's atmosphere by computing the connectivity of the magnetic field. To do so, different techniques have been investigated. For example, bright coronal loops appearing in hot temperature filters can be interpreted as consequences of reconnection, or exchange of flux in a newly formed domain, happening in a nearby region. Knowing how reconnection occurs in the dissipation regions related with those topological/geometrical features also informs on the reconnection regimes taking place (see section{slipping}), as well as indicate observational predictions that can be fulfilled with higher resolution observation.

\subsection{QSLs and HFTs}  %%%%%%%%%%%%%%%%%%%%%%%%%%%%%
\label{sect_QSLs}

%    {\S\bf History} \\

In \citet{Demoulin1994b}, the authors computed the topology of several flaring regions and looked at the locations of nulls, obtained both with potential and linear force-free field models. These locations were then compared with that of energy release seen in UV and X-rays. They then found that the location of the null, or even their presence, was not necessary to explain the different locations of energy release, whereas the spatial properties of the coronal field was. 

A few years earlier, \citet{Hesse1988} and \citet{Schindler1988}  had also proposed that in a 3D volume, magnetic field dissipation should include all effects with localised non-idealness that leads to a parallel component of the electric field along the magnetic field. Then, reconnection does not need to be associated with magnetic nulls, closed field lines or other singular structures. For magnetic reconnection to occur, there is a need to create a dissipation layer where the electric current density is important: \citet{Priest1995} and \citet{Demoulin1996b} added that such regions are created if the magnetic field connectivity is strongly distorted but can still remain continuous.

In \citet{Aulanier2005} and \citet{Pontin2005}, the authors investigated the evolution of the magnetic field in a 3D MHD simulation in the absence of magnetic null points, but with the presence of a nonideal region formed by shearing/twisting boundary motions. They noted that the field lines would change their connections continually and continuously in the volume with strong current concentrations. They verified, with these simulations, that in the presence of a strong $E_{\parallel}$ component, reconnection could still happen without the presence of null points.

%    {\S\bf Concept of QSLs - mapping norm} \\

In the mathematical sense, null points are strictly defined as the locations where the magnetic field vanishes, and are accompanied by separatrices and separators. However, {a broader class of structures may be defined: these are quasi-separatrix layers (QSLs). When thin enough, QSLs behave physically as separatrices even though there are no true mathematical discontinuities of the field line mapping.}
% Strong currents generally develop at QSLs during an evolution of the magnetic configuration, for example due to photospheric motions.

\begin{figure}  %________________________ FIG ______________________________________    
\centering
\includegraphics[width=1\textwidth,clip]{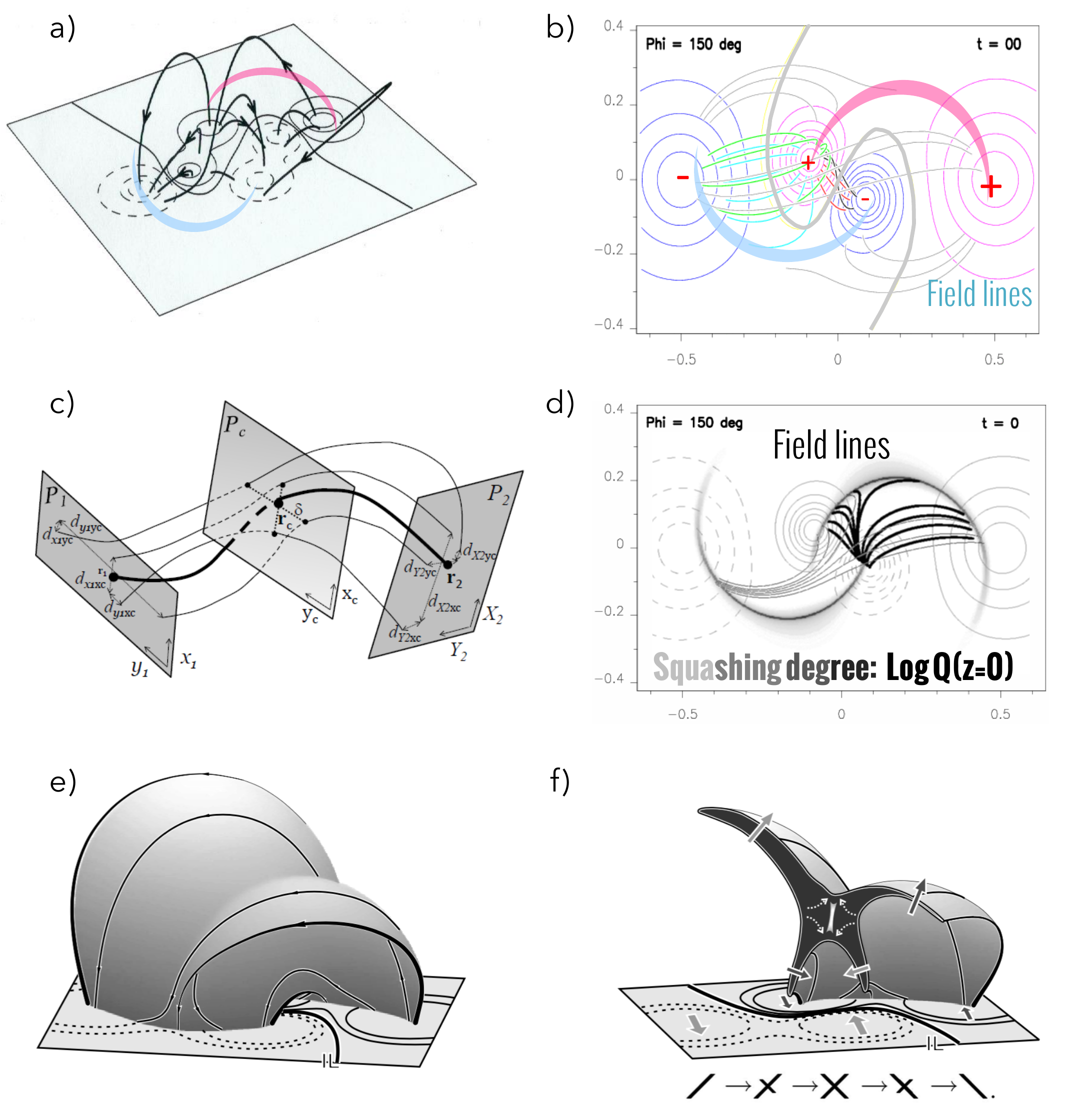}
\caption{(a) A set of magnetic field lines randomly traced in a quadrupolar configuration without a null point. The trace on the lower plane of the largest gradient of connectivity, \ie\ the QSLs, are shown with pale blue and magenta crescent areas. They are located within magnetic polarities (dotted and plain isocontours on the surface) of the same sign. (b) Quadrupolar configuration analysed in a numerical setup by \cite{Aulanier2005} where field lines are traced in different colors depending on their anchoring region. For example, the green and blue set of field lines are departing from the same positive polarity (in magenta), but are seen to connect to the different negative polarities (blue isocontours). As such, one can trace the connectivity gradient region (in magenta). (c) The field line mapping and the squashing degree $Q$ can be calculated following the technique of \citet{Pariat2012}, which is illustrated here by a generic connectivity between two local planes while the QSL trace is computed on the central plane. (d) The QSLs are computed numerically: their traces on the photospheric plane are shown in gradient of grey, with the darker greys indicating higher values of the squashing degree $Q$. Two sets of field lines are added with their footpoints selected on a segment crossing the QSL trace. They show a divergence pattern characteristic of field lines accross QSLs. The whole QSL volume is represented in perspective in (e), for a similar quadrupolar configuration. (f) A cut within the volume  shows the X-shaped morphology of the QSLs, also called a Hyperbolic Flux Tube \citep[adapted from][]{Titov2002}.
}
\label{fig_4}
\end{figure}

This generalisation to QSLs was introduced by the seminal works of \citet{Priest1995} and \citet{Demoulin1996}. QSLs are defined as regions of space where the connectivity of the magnetic field is drastically changing, while remaining continuous in general. This is for example illustrated in \fig{4}a where two bipoles are shown. The pale blue and magenta crescent regions indicate the footpoints of QSLs in a schematic way. The configuration does not have a null point but as the flux concentrates in each magnetic poles, it still creates a gradient in the connectivity of the field lines.

To properly define regions of strong changes of connectivity, there is a need to introduce the mapping of field lines, which follows the connectivity of a field line from one footpoint to another in a given magnetic configuration. In the solar context, this lower boundary is introduced because of the line-tying of the field lines in the dense and slowly evolving photosphere. This boundary is in general not required to define QSLs \citep{Demoulin1996b}. Below, we keep this description because of its direct solar application. For that, the mapping norm was introduced \citep[Eq.4 in][]{Demoulin1996b}:

\BE
N= \sqrt{ \bigg(\frac{\partial X}{\partial x}^2+\frac{\partial X}{\partial y}^2+\frac{\partial Y}{\partial x}^2+\frac{\partial Y}{\partial y}^2 \bigg)}
\EE

where $(x,y)$ represent one footpoint location at one plane (\eg\ see \fig{4}), and $(X,Y)$ the corresponding other footpoint of the same field line. A QSL is defined as any region where $N\gg 1$.
A strongly distorted connectivity region, or QSL, means that when tracing magnetic field lines anchored in the same vicinity, their opposite footpoints will be located at very different positions in the opposite polarity (\fig{4}b,d).

%    {\S\bf Squashing factor} \\

However, the mapping norm is only defined by a neighbouring field line region in one polarity: its values therefore depend on which polarity the field lines have been chosen (it is footpoint-dependent). To estimate a non-footpoint dependent parameter, \citet{Titov2002} defined another parameter, the squashing factor $Q$. It is independent of the footpoint as it is weighed by the ratio of the vertical field component at each footpoint. Similarly to $N$, $Q$ provides an information on the distortion of the magnetic field connectivity. By introducing the two footpoint mappings $N_{12}$ and $N_{21}$ :

\BA
N_{12}= \sqrt{ \bigg(\frac{\partial X_2}{\partial x_1}^2+\frac{\partial X_2}{\partial y_1}^2+\frac{\partial Y_2}{\partial x_1}^2+\frac{\partial Y_2}{\partial y_1}^2 \bigg)}\\
N_{21}= \sqrt{ \bigg(\frac{\partial X_1}{\partial x_2}^2+\frac{\partial X_1}{\partial y_2}^2+\frac{\partial Y_1}{\partial x_2}^2+\frac{\partial Y_1}{\partial y_2}^2 \bigg)}, \label{eq_mappingnorm}
\EA

the squashing degree was defined as: 

\BA
Q=Q_{12}= \frac{N^2_{12}}{|B_{z,1}(x_1,y_1)/B_{Z,2}(X_2,Y_2)|}\\
=Q_{21}= \frac{N^2_{21}}{|B_{z,2}(x_2,y_2)/B_{Z,1}(X_1,Y_1)|} .\\
\EA

Then, QSLs represent the regions with the highest squashing factor $Q$. Recently, a computational method to obtain the squashing degree within a 3D domain was proposed by \citet{Pariat2012}  (see \fig{4}c). This is further developed in \citet{Tassev2016}.
%    {\S\bf HFT} \\

Since the QSLs form a thin volume (\eg\ \fig{4}e), one can define the region where the field lines diverge the most in the central part of this volume. In \fig{4}f, a transverse cut in the volume is shown, which displays 4 branches marking the characteristic shape of QSLs. This particular region was coined a hyperbolic flux tube \citep[HFT, see][]{Titov2002}. HFTs can also be understood as the ``intersection'' of two QSLs, and as such are generalising the concept of a separator line in 3D in the context of QSLs (\ie\ {since a separator is formed by the intersection of two separatrices}). Their shape is particularly reminiscent of an X-point in the transverse cut. At the bottom of \fig{4}f a schema indicates its shapes following the location of the cut in the volume (\ie\ those shapes are obtained by doing different cuts of the HFT orthogonal to the local magnetic field). Closer to the anchoring surface, the HFT ressembles a line: they are the inner regions within the pale blue and pale magenta traces shown in \fig{4}b. Further away from the photosphere, the HFT displays an X-shaped structure. {Recently, the concept of the squashing factor to define QSLs and HFTs has been extended to ``slip-squashing'' factors \citep{Titov2009} (from the discussion in \citet{Hesse2005}), which allows the characterisation of the change in the magnetic field connections.}

\subsection{QSLs in the presence of flux ropes: from theory to observations}  %%%%%%%%%%%%%%%%%%%%%%%%%%%%%
\label{sect_FRQSLs}

%    {\S\bf QSLs in the presence of flux ropes} \\

\begin{figure}  %________________________ FIG ______________________________________    
\centering
\includegraphics[width=1\textwidth,clip]{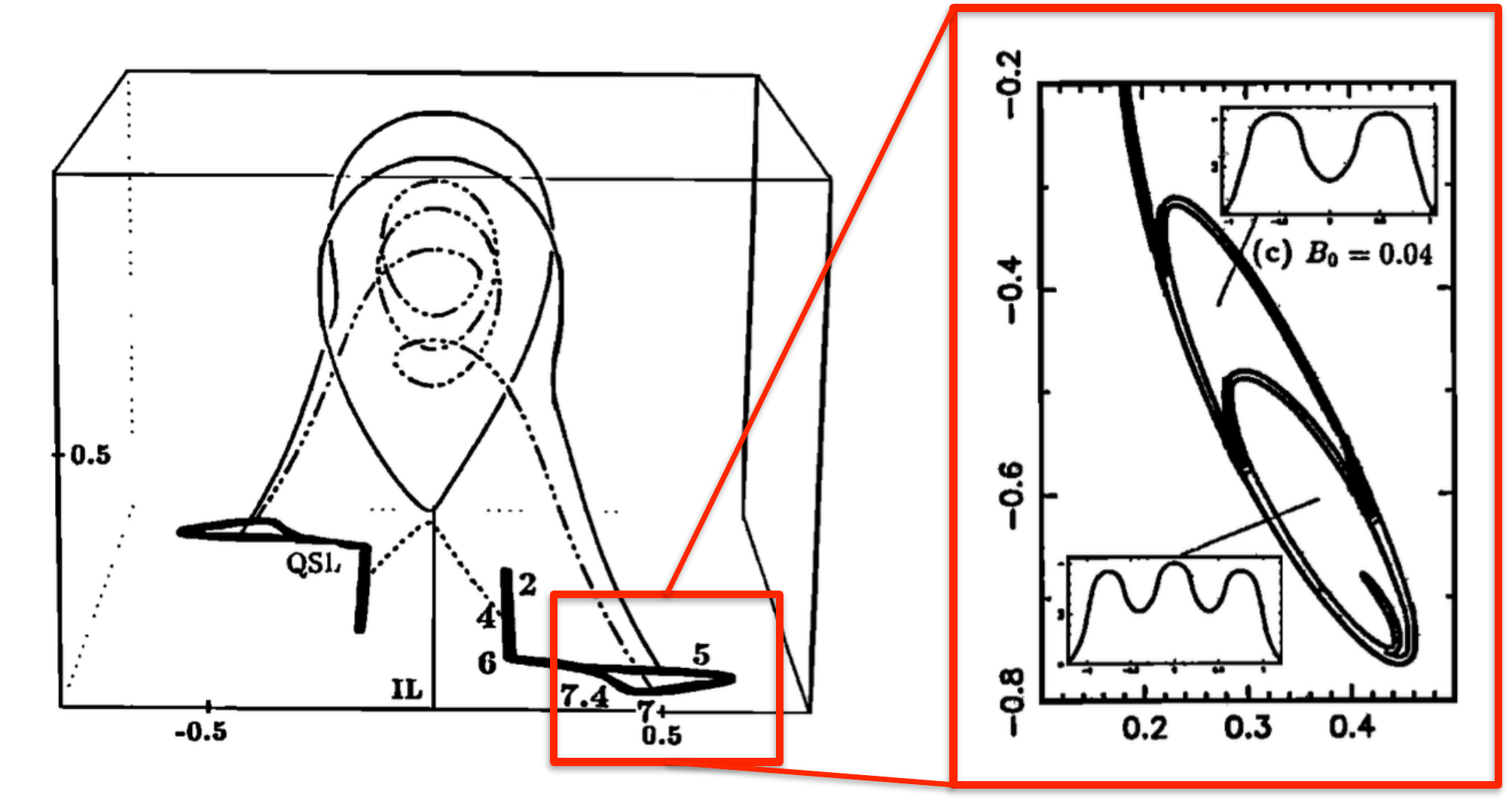}
\caption{Projective view of a configuration containing a flux rope, as indicated with the dashed-dotted (three turns) and solid (one turn) twisted field lines. The small, dotted field line represent a coronal loop lying underneath the flux rope. The gradient of connectivity between those field lines is indicated with the elongated, bold lines at the photospheric level (QSL trace). Their straight part is associated with the low-lying coronal loop, while the round region is associated with the anchoring region of the twisted field lines. A zoom in the region shows a hook-shaped of the flux rope anchoring region, where a higher twist corresponds to a higher number of swirls \citep[adapted from][]{Demoulin1996}.
}
\label{fig_5}
\end{figure}

In the presence of a twisted magnetic field, the quasi-separatrix layers trace the frontier between the twisted magnetic field and the more potential surrounding field (see \fig{5}). Their traces on the lower boundary show a typical \J-shape, with a straight part associated with low-lying coronal loops while the hook region of the \J\ shape is associated with the anchoring region of the flux rope. In \citet{Demoulin1996}, the authors investigated the morphology of the QSLs and especially the hook region: because the QSLs form a thin volume, that delimitates the frontier between differently connected field lines, this volume is also connected to the photospheric surface. Then, the more the magnetic field is twisted, the more complex the QSL photospheric footprint becomes: the QSL swirls around the flux rope as the twist of the structure increases (\fig{5}b). QSLs in the presence of flux ropes were also investigated in laboratory experiments. \citet{Gekelman2012} provided experimental evidence that QSLs indeed form in the presence of flux ropes created in a laboratory device.
\\

%    {\S\bf Currents at QSLs} \\

\begin{figure}  %________________________ FIG ______________________________________    
\centering
\includegraphics[width=1\textwidth,clip]{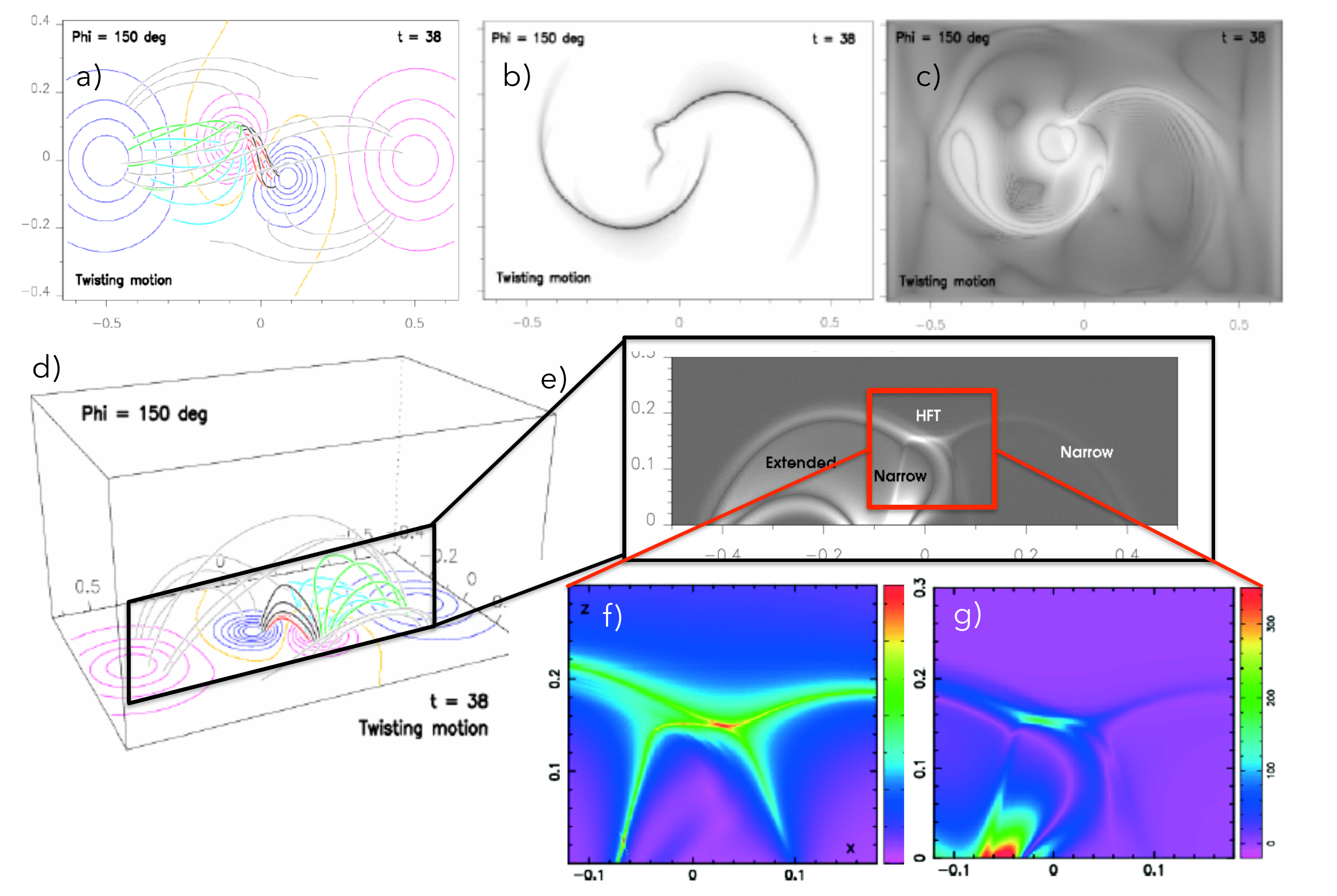}
\caption{(a) Top view of a quadrupolar magnetic configuration (with two bipoles, where the positive (resp. negative) polarity is indicated in magenta (resp. blue)). A photospheric velocity field is applied as a boundary condition so as to reproduce a twisting motion in the small positive polarity. The photospheric traces of the associated QSLs are shown in (b): the highest values of the squashing degree $Q$ are shown in black. The electric currents are shown in grayscale in (c), with the most intense currents shown in white. (d) Side view of the configuration, with (e) showing a transverse cut in the middle of the domain of the coronal current density. The strongest currents are seen to appear at the locations with the highest squashing degree or HFT, as is also shown in the color-coded zooms of the QSLs (f) and the currents (g). \citep[adapted from][]{Aulanier2005}.
}
\label{fig_6}
\end{figure}

Because of the strong distortion of the magnetic field line mapping at QSLs, the latter were proposed as preferential locations for current buildup \citep{Demoulin1996b,Demoulin2005}. This was investigated in numerical simulations, and several authors indeed showed the generic formation of strong current density regions at QSLs, as indicated in \fig{6} \citep{Aulanier2005, Pariat2006, Masson2009, Effenberger2011, Craig2014}.
% In particular, \citet{Titov2003,Galsgaard2003} showed that photospheric twisting motions were very efficient in forming layers of strong current density within HFTs. 
However, an exact one-to-one correspondence between the localisation of the highest squashing degree values (HFT) and the highest current densities are not necessarily found, as was shown in \citet{WilmotSmith2009} and \citep{Janvier2013}. The correspondence of QSLs with regions of reconnection is not limited to MHD models generally investigated in the context of the Sun. Recent kinetic simulations in 3D such as by \citet{Wendel2013} have shown that QSLs are associated with areas of large gradients of parallel electric field, which provide a new understanding on the determination of reconnection sites in dominantly collisionless plasmas such as the Earth's magnetosphere. Since electric currents are extremely difficult to directly investigate in the solar corona (because of the absence of reliable direct observations of magnetic fields in the Sun's atmosphere), simulations provide a useful environment to understand the formation and the evolution of currents in the presence of QSLs, such as shown in \fig{6}. Recently, proxies of coronal currents, such as their photospheric signatures, have been used to investigate the similarities in the location, morphology and evolution of QSLs and currents. They show the correspondence expected from numerical simulations, as detailed in \sect{obscurrents}.

 %   {\S\bf Solar QSLs} \\

Since QSLs generalize the concept of separatrices and are associated with locations of electric currents, and hence reconnection, investigation of the topology of flaring magnetic field configurations naturally led to the  search for QSLs and their relation with flares. For example, \citet{Demoulin1996b, Demoulin1997} used modelling techniques of the coronal field to reproduce magnetograms of flaring regions and computed QSL locations. In these works, the authors found that locations where H$\alpha$ brightenings were observed were related to locations of QSLs. Furthermore, QSLs can also be found in the presence of a null point: such a configuration was also investigated in a circular ribbon flare, using data-driven simulations \citep{Masson2009}.
Since then, investigating the locations of QSLs has proved to be successful at interpreting a large variety of flaring regions \citep[\eg][]{Schmieder1997, Bagala2000,Mandrini2006,Restante2009,Chandra2011}.\\

%    {\S\bf QSLs in eruptive flares} \\

From then on, the locations and the morphology of sudden flare brightenings were compared with the locations of QSLs found with a magnetic field model. Over the years, the techniques refined, helped by higher spatial resolution instruments as well as more refined modelling techniques (such as magnetic field extrapolations). To give an example, the ribbons of eruptive flares were often compared with QSLs found in the presence of flux ropes: indeed, this type of flares often displays the presence of two flare ribbons appearing in H$\alpha$ \citep{Chandra2009}, and in particular these ribbons often display a \Jshape-shape structure. This morphology is very similar to the analytical shapes found for QSLs in the presence of twisted flux tubes (see above and \fig{5}), as well as in recent 3D MHD numerical simulations of eruptive flares \citep[see Fig. 3 in][]{Janvier2013}. 

\begin{figure}  %________________________ FIG ______________________________________    
\centering
\includegraphics[width=0.8\textwidth,clip]{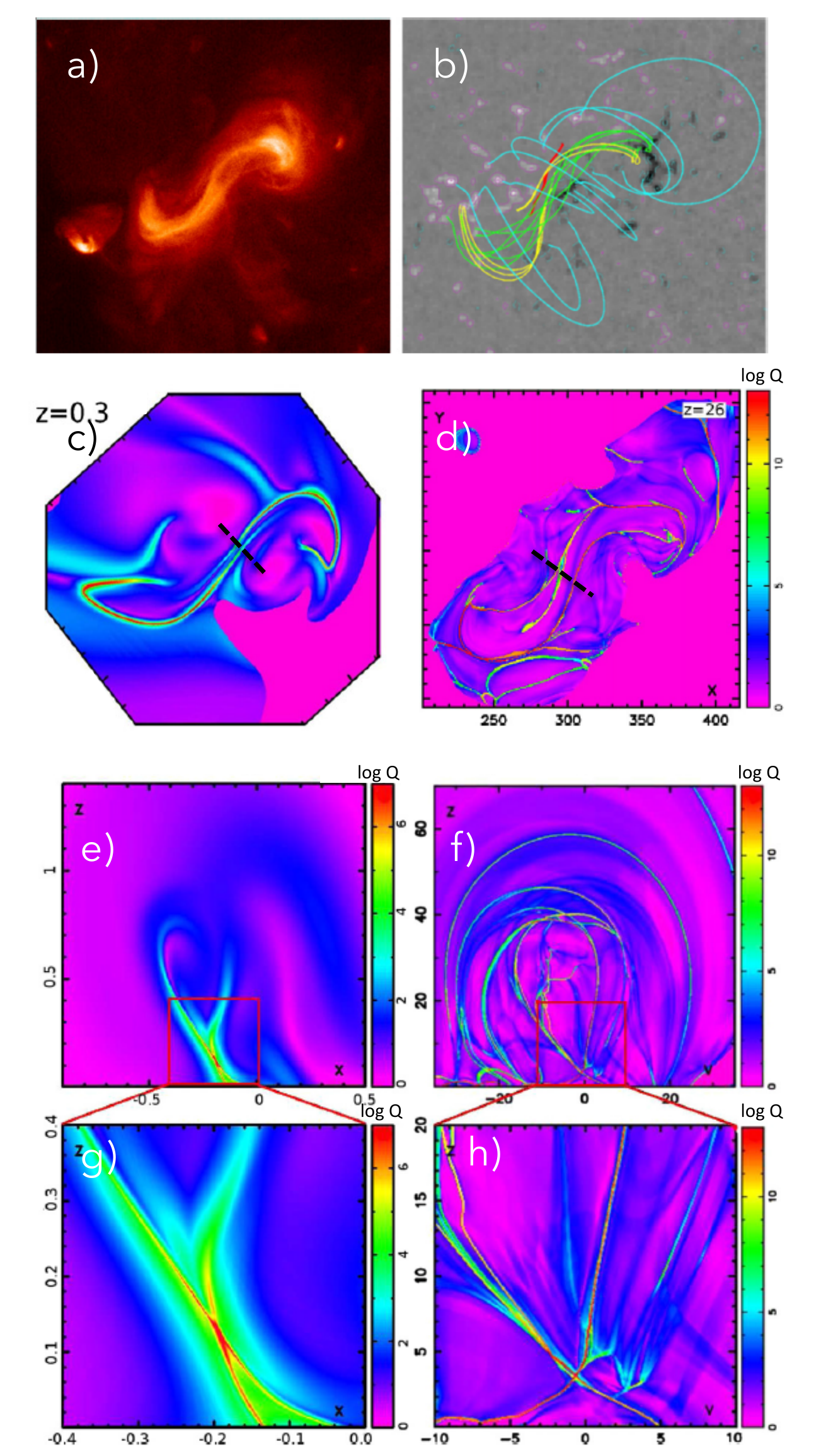}
\caption{(a) Sigmoidal region seen by the XRT instrument aboard Hinode from which a magnetic model is constructed, shown with a sample of field lines in (b). Panels (c,e,g) show the numerical simulation of an unstable flux rope, while panels (d,f,h) show similar plots for a magnetic configuration derived from observations (panel b). The (near) photospheric traces of the QSLs for a numerical flux rope simplified model are shown in (c). They are compared in panel (d) with that of the configuration created by a flux rope inserted in the extrapolated potential magnetic field of the magnetogram shown in (b). They both display the typical \Jshape-shape expected in the presence of a flux rope. A transverse cut (dashed black lines in c and d) is shown in the magnetic field numerical model (e) and the extrapolated magnetic field (f), where the location of the highest values of the squashing degree $Q$ is found underneath the flux rope in both cases. A zoom indicate the presence of a HFT in (g,h) \citep[adapted from][]{Savcheva2012}.
}
\label{fig_7}
\end{figure}

\citet{Savcheva2012} investigated the shape of QSLs for a magnetic field model of an active region which was associated with the presence of a sigmoid \citep[a region where the coronal loops display $J$ or $S$-shaped features, indicative of a flux rope, see \eg][]{Green2007,McKenzie2008}. They model the flux rope associated with the sigmoidal region using the flux rope insertion method \citep[see][]{VanBallegooijen2004}. With a magnetic model of the active region, the authors could investigate the connectivity change in the volume of interest, and look in more details at the morphology of the QSLs. Some of these results are reproduced in \fig{7}, where it shows that the photospheric traces of the QSLs are much more complex than in analytical and simulation models, yet still displaying a hooked, \Jshape-shape. A transverse cut in the middle of the twisted structure also shows the presence of an $X$-shaped region that is reminiscent of the HFT (\fig{7},e-h). Such a work therefore provides observational evidence that HFTs and associated QSLs are generic features found in the presence of sigmoidal regions and flux ropes.\\

 %   {\S\bf Recent extensions} \\

Recently, such an analysis has been extended to a variety of flaring regions and with different techniques. For example, \citet{Zhao2016} studied a flaring region and applied another technique to reconstruct the magnetic field of the region, namely, a non-linear force-free field model \citep[following the technique of][]{Gilchrist2013}. They found similar morphologies in the shape of the QSLs as in \citet{Savcheva2012}. As such, the morphology of QSLs in the presence of erupting flux ropes is independent of the extrapolation method. This is expected because of the structural stability of QSLs, in contrast to the presence and the location of null points and associated separatrices, \citep[see][on the influence of the extrapolations on the stability of null points and QSLs]{Demoulin1994b, Janvier2016}. Furthermore, the shapes of the QSLs can be directly compared with that of the flare ribbons.

Then, evolving the magnetic model of an active region, for example by an MHD evolution or a magnetofrictional relaxation (in the case of an unstable flux rope) also allows the investigation of the evolution of QSLs. The multiple time sequences of these evolutions can provide snapshots that can be compared with observations of flare ribbons, as presented in \citet{Savcheva2015,Savcheva2016,Janvier2016}.

As such, the search for QSLs in a magnetic field reveals the coronal locations of the dissipation, as well as a tool to interpret the photospheric signatures of reconnection as seen with flare ribbons.

%%%%%%%%%%%%%%%%%%%%%%%%%%%%%%%%%%%%%%%%%%%%%%%%%%%%%%%%%%%%%%%%%%%%%%%%%%%%%%%%%%%%%
\section{Dissipation layer} %%%%%%%%%%%%%%%%%%%%%%%%%%%%%
\label{sect_currents}

\subsection{Dissipation process}  %%%%%%%%%%%%%%%%%%%%%%%%%%%%%
\label{sect_dissipationprocess}

%    {\S\bf Resistive current layers} \\

The first developments in the theory of magnetic field reconnection were accompanied by a description of the dissipation process taking place in current layers. The diffusion terms appear in Ohm's law as non-ideal terms. In its simplest description (resistive MHD for a single fluid), the non-ideal departure is described with $\eta \vec{J}$:
\BE
\vec{E}+\vec{v}\times \vec{B} = \eta \vec{J} \label{eq_resistivelaw}
\EE
where $\eta$ is the plasma resistivity (a measure of the collisionality of the plasma, see \citet{Spitzer1956}) and $\vec{J}$ is the electric current density vector.
The effect of finite conductivity on the diffusion of magnetic fields was already investigated for several configurations by \citet{Parker1956}. However, simply considering the diffusion of magnetic fields in the corona provides timescales (years) that are considerably larger than that of flares (minutes).
%Why doesn't the Sweet and Parker model work? The solar corona is a very good conductor: the magnetic Reynolds number (or Lundquist number) is expressed with the parameter $S = 4 \pi \sigma v_A L/ c2 \approx 10^{10-12}$. In the expression, $v_A$ represents the Alfv\'en speed, $L$ is a characteristic global length scale, $\sigma$ is the plasma conductivity, and $c$ is the speed of light. Then, considering an inflow velocity $v_in$, and an outflow velocity of the same order as $v_A$, the reconnection rate, given as the ratio between the outflow and the inflow, gives: $v_in/v_A=S^{-1/2}$. This gives an inflow of the order of $10^{-5\sim6}v_A$. This is not very fast. Then, the Petschek reconnection model proposed speeds of the order of $v_in \sim 0.10 v_A$.\\
Then, \citet{Petschek1964} introduced another reconnection model where he instead considered a much shorter current sheet, which provided a much greater inflow speed needed for fast reconnection. In his model, he overcame the elongation of the current layers (from an X shape to a double Y shape) by invoking the appearance of oblique, magnetohydrodynamic, standing shock waves which develop at the edges of the current layer. Then, the reconnection rate is enhanced by hydromagnetic actions (since the waves allow magnetic energy conversion) outside the diffusion region, while the ohmic dissipation only occurs in the very small current layer. However, its mechanism was doubted \citep{Green1967} and later on proved to not be reproducible in numerical simulations either \citep{Biskamp1986}{, unless when some variation in the resistivity was accounted for \citep[see][]{Baty2009}. Further developments of the Sweet-Parker and Petschek models were compared, such as by \citet{Kulsrud2001}, who investigated analytically the similarities between the Sweet-Parker and the Petschek models. Both are found to have a similar reconnection rate in case of constant resistivity due to the imposed condition that the length of Petschek's diffusive layer is not a free parameter, a condition that was not taken into account in the original work. The Sweet-Parker current sheet, which describes a stationary model, is quite restrictive compared with the possible evolutions of current sheets, as }described below in \sect{currentlayer}. {We also refer the reader to the historical evolution and recent advances deriving from the Sweet-Parker configuration discussed in \citet{Loureiro2016}}.

%  {\S\bf Ambipolar diffusion} \\

A more complete description of the physical dissipation mechanism in current layers can include ambipolar diffusion, which is effective in partially ionised gases of sufficient density \citep{Piddington1954b}. This process is related to the drift of the plasma with respect to the neutrals in the presence of a magnetic field \citep[see][for more insights on the phenomenon and its importance in astrophysical systems]{Zweibel2015}. The ambipolar diffusion does not itself lead to magnetic reconnection, but as it squeezes the magnetic field lines together, they end up ``piling up'', permitting a more efficient dissipation \citep[see also][]{Parker1963}. Ambipolar diffusion has recently been revisited in the context of reconnection in the chromosphere, where the plasma is dense \citep[see \eg][]{Leake2013}. However, it is dubious that ambipolar diffusion may have a strong role in the corona since the density there is rather low and the corona is fully ionised.

%  {\S\bf Turbulence} \\

The effects of turbulence on reducing the plasma conductivity were already pointed out by \citet{Sweet1950}. \citet{Wentzel1963} proposed also early on that the increased dissipation due to turbulence can increase the magnetic annihilation rate. Plasma experiments such as those conducted by \citet{Baum1976} then showed direct observations of rapid magnetic field reconnection, and transition from slow to fast reconnection due to an anomalous resistivity, validating such reconnection models. The effect of anomalous resistivity was studied numerically in different current sheet configurations: it was for example shown to account for the transition from a Sweet-Parker-like current layer to a Petscheck type current layer \citep{Baty2009,Baty2014}. The effects of stochastic magnetic fields  were also shown to increase the reconnection rate \citep[\eg][]{Lazarian1999}.\\

%  {\S\bf Small-scale} \\
  
 A large body of work has investigated the reconnection mechanism at small scales. Effects such as the Hall term \citep{Piddington1954a}, the ion inertia term and the electron pressure density can play a role in the reconnection mechanism and can change the energy partition. The relevant Ohm's law in such a case is as follows:
 \BE
\vec{E}+\vec{v}\times \vec{B} = \eta \vec{J} + \frac{1}{en}\vec{J}\times\vec{B} - \frac{1}{en} \nabla\cdot\vec{P}_e - \frac{m_e}{e}\frac{d\vec{v}e}{dt}
\label{eq_generalisedlaw}
\EE
where the first term (on the right hand side of the equation) is the plasma resistivity term, the second the Hall term, the third the electron pressure tensor, and the remaining term represents the electron inertia.

\begin{figure}  %________________________ FIG ______________________________________    
\centering
\includegraphics[width=0.8\textwidth,clip]{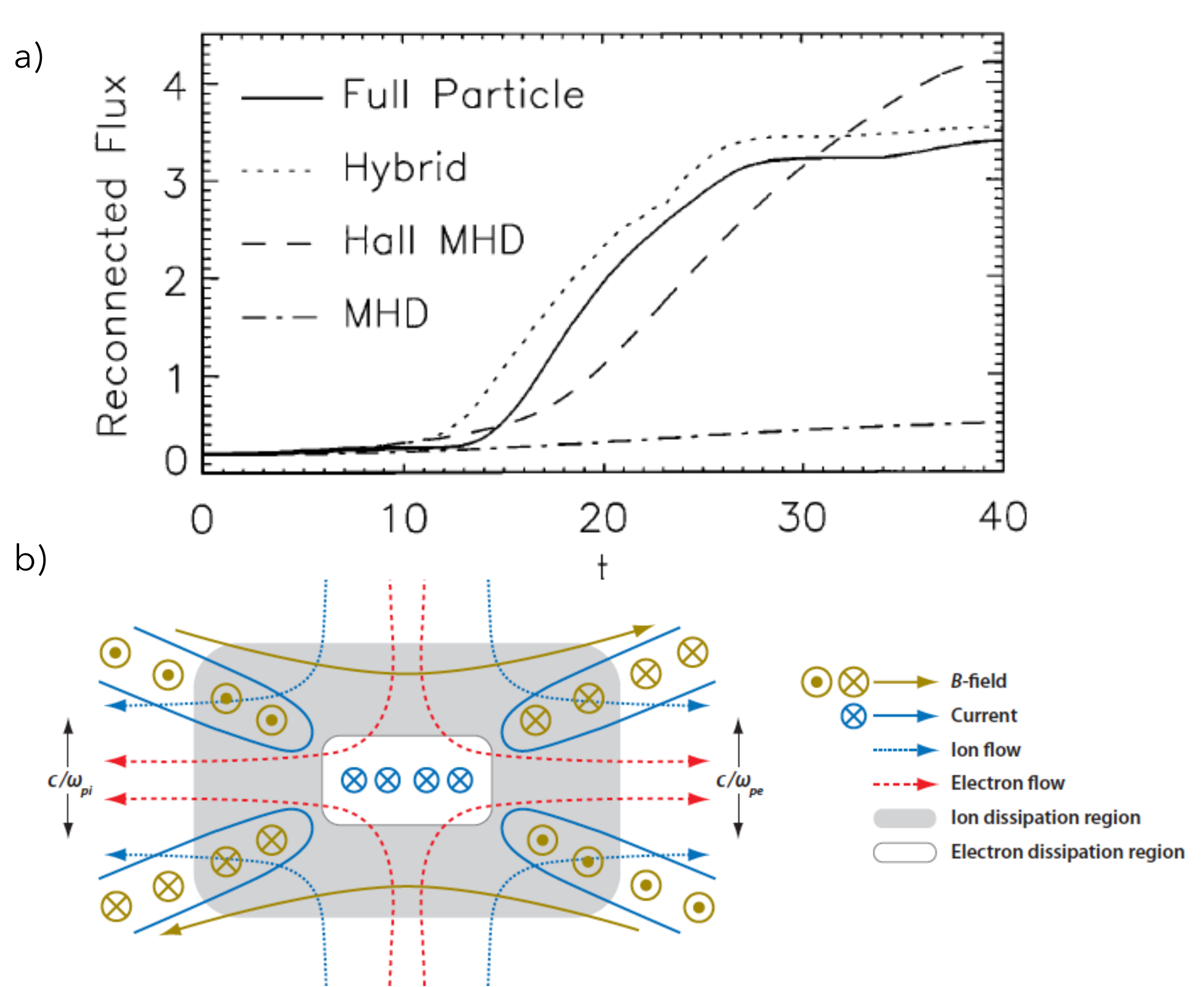}
\caption{Comparison of resistive MHD and kinetic descriptions of the current layer. (a) Results of the Geospace Environmental Modeling (GEM) magnetic reconnection challenge, where several codes (MHD and PIC) were tested to investigate the effects of the nonlinear terms described in the generalised Ohm's law. It was found that codes that include the Hall term did not differ much one from another, while a conventional resistive MHD description of reconnection did not agree with all the other results. Here, the time evolution of the reconnected flux in those simulations are shown to indicate the differences \citep[adapted from][]{Birn2001a}. (b) Description of the magnetic field geometry in collisionless reconnection, where the flows of the ions and the electrons are decoupled in the diffusion area \citep[adapted from][]{Zweibel2009}.
}
\label{fig_8}
\end{figure}

These last three terms arise where the traditional MHD formalism breaks down for a current sheet that is thin enough, \ie\ the diffusion region becomes thinner than the ion skin depth (in such a case, we generally refer to reconnection as being collisionless). Note that adding more terms to the Ohm's law means that the physics related to the ions and the electrons is more complete than a description where the plasma is considered as a single species fluid, as can be seen by comparing \eq{resistivelaw} and \eq{generalisedlaw}. For example, the Hall term arises when magnetic drifts are considered: this relates to the motion separation between electrons and ions, \ie\ electric currents, which then need to be added as a correction to Ohm's law \citep[the $(ne)^{-1}\vec{J}\times\vec{B}$ term, see numerical developments in][]{Huba2003}. We note that a careful interpretation of the Hall term is needed, since the term itself is not a magnetic field connectivity breaking term. Indeed, the Hall term decouples the ions and the electrons, and the magnetic field remains tied to the electrons (meaning that at the electron-scale, the magnetic field still remains ``frozen-in''). The importance of small scales effects was pointed out by \citet{Birn2001a}, where the authors investigated the effects of different numerical setups with different expressions of the Ohm's law. In particular, any implementation (MHD or PIC) containing the Hall term gave similar results, contrary to the simple resistive MHD description (see \fig{8}a). 
The decoupling of the ions and the electrons are shown in \fig{8}b, where the diffusion region is defined differently (with different physical effects) for the ions and the electrons.
Studies of small-scale effects therefore include Hall fields \citep{Drake2008}, anisotropic pressure \citep{Birn2001b}, electron dynamics and viscosity \citep{Ma1996,Hesse2004,Cai2008}, while reviews of these effects can also be found in \citet{Porcelli2002,Shay2004,Buchner2007}.

\subsection{Current layer formation and evolution}  %%%%%%%%%%%%%%%%%%%%%%%%%%%%%
\label{sect_currentlayer}

%  {\S\bf Steady-state/Driven reconnection} \\
  
In the Sweet and Parker's mechanism \citep{Sweet1956, Parker1957b,Sweet1958a}, and as well as in the Petschek mechanism \citep{Petschek1964}, reconnection is considered analytically in a steady-state current layer, treated as a boundary layer where the plasma outside of this region is considered as current-free and perfectly conducting. Such models are useful in obtaining the magnetic energy conversion (or reconnection) rates \citep{Parker1963}, as well as providing the geometrical characteristics of the current layer \citep{Priest1975,Tur1976}. However, in the works of \citet{Green1967} and \citet{Petschek1967}, the authors had to modify the original Petschek reconnection scenario (\eg\ with additional waves) as they recognised that the region where ohmic diffusion takes place at the intersection of two shocks is too thin to be hydrodynamic. \citet{Friedman1968} also pointed out that the current density in Petschek's original model violates the two-stream instability criterion. Finally, Syrovatskii's work \citep{Syrovatskii1966} on dynamic dissipation in current layers throws some doubt on whether a quasi-steady state can be achieved at all. 

It is necessary to remind the reader that reconnection in such steady-state configurations is driven by boundary motions: the magnetic field is forced to move into the dissipation layer by a driving inflow around the reconnecting region \citep{Hahm1985,Wang1996}. The rate of reconnection or energy conversion is therefore dictated by the inflow of magnetic flux brought to the dissipation region. The dynamics of coronal field can infer flows which velocity fields drive/enhance the reconnection in the current layer. Therefore, a Sweet-Parker regime is often a misnomer when describing reconnection processes occuring in the Sun's corona, since the inflows from the surroundings of the reconnection region are not themselves steady-state. {How the reconnection rate is actually determined by boundary conditions is still poorly understood \citep[see the discussions in][]{Comisso2016}. Recently,  \citet{Forbes2013,Baty2014} have argued that the rate of reconnection is more likely to be controlled by the nonuniformity of the normal magnetic field in reconnecting layers rather than by the boundary conditions.}
%Recently, simulations of driven reconnection in collisional or collisionless plasmas \citep{Horiuchi2001, Ishizawa2001} have shown that driven reconnection are reponsible for the formation of closed magnetic flux areas/volumes (so-called magnetic islands or flux rope in 3D). Its consequences in the light of solar flares are summarised in \sect{plasmoids}.

Steady-state, driven reconnection models lose the description needed to explain the onset of flares. Indeed, as was already recognised by \citet{Gold1960}, it is important for a model to include the suddeness of flaring events. In particular, they acknowledged the necessity of force-free fields to progressively store free magnetic energy, which exist prior to the flare. They also recognised the need to have a storage mechanism and release mechanism, invoking the existence of instabilities.\\

%    {\S\bf Tearing Instabilities} \\
    
The instability of current layers gained some attention in the early stages of plasma laboratory experiments, especially because they were found to be detrimental to the realisation of fusion, the process to harness a star-like energy. Then, \citet{Furth1963} proposed that a resistive instability takes place in the current layer, which leads to a reajustment of the magnetic field in a configuration of lower energy. This lower state is characterised by the presence of magnetic islands (2D transverse cuts of 3D twisted flux ropes, appearing as nested magnetic field lines). As it gives a shredded structure to the current layer, this instability was coined the tearing instability. Furthermore, the interest in this mechanism for solar flare application lies in the fact that the timescale of the tearing instability was much smaller than a simple resistive diffusion of the equilibrium (although the instability grows more slowly than MHD instabilities, as $t_{A} < t_{T} < t_{R}$ where $t_{A}$ is the MHD time-scale (Alfv\'en time), $t_{T}$ the tearing instability time scale and $t_{R}$ the resistive diffusion time scaling with $\eta^{-1}$). Putting numbers for the solar corona, \citet{Jaggi1963,Sturrock1968,Kliem1995} found that the e-folding time of the instability was small enough to account for the time scale generally associated with solar flares {\citep[although this is only valid in the case of extremely small scales, see discussions in][]{DelZanna2016}}, while \cite{Spicer1981} put forward this instability as a possible mechanism for particle acceleration. These findings were nonetheless challenged by the incapacity of a single tearing layer to provide the necessary current dissipation to account for the thermal emissions seen in flares \citep[\eg][]{LaRosa1990}. A more complex system of small-scale current layers (or multiple tearing layers) was proposed instead. 2.5D simulations (\ie\ with an invariant direction) of current sheets in the context of solar flares were also recently investigated \citep{Shibata2001,Lynch2016}, where the breaking up of the current layer in multiple plasmoids was shown. {The connection between plasmoids investigated in numerical simulations, whether 2D or 3D, and observations of solar flares, is discussed in more details in \sect{plasmoids}.}

A variety of resistive instabilities have since then been studied, such as the resistive kink \citep[with mode number $m=1$, also coined fast tearing as its growth rate scales with $\eta^{1/3}$, see \eg][]{Rosenbluth1973,Hazeltine1986,Waelbroeck1989,Watanabe1995}. {Although the resistive kink instability has been extensively studied in the context of tokamak plasmas, kink modes (resistive in nature) can be put forward to explain the onset and subsequent magnetic energy dissipation of eruptive flares. This was discussed in \citet{Hood1979} for line-tied field lines (relevant in the context of solar flares), and  numerical simulations of kink unstable flux ropes were compared with observations in \citet{Torok2005}.}  
Furthermore, multiple tearing instabilities can greatly enhance the reconnection rate in the nonlinear regime due to coupling between islands \citep[where the reconnection outflow of a magnetic island can increase the inflow of a nearby reconnection site, \eg][]{Pritchett1980, Ishii2002, Bierwage2005, Wang2007, Janvier2011, Wang2015}. Finally the effect of line-tying, relevant in the context of coronal magnetic field, was investigated in \citet{DelZanno2008}, which found that the growth rate scaled with $\eta$ for small structure lengths (compared with the system size), while was tearing-like for long structures. This may be of interest considering different reconnecting coronal loops in the context of say, nanoflares involving small loops versus the larger structures generally seen in flares. It should also be noted that recently, the tearing mode has been revisited {in the context of current sheets with large aspect ratios. In such cases, it is possible to find current layers for which the tearing mode growth rate is of order unity and independent of the Reynolds number: such modes are coined ``ideal tearing modes" and are interesting to consider in the context of solar flares} \citep{Pucci2014,Landi2015,DelSarto2016,DelZanna2016}. \\

 % {\S\bf Current layers in 3D} \\

\begin{figure}  %________________________ FIG ______________________________________    
\centering
\includegraphics[width=1\textwidth,clip]{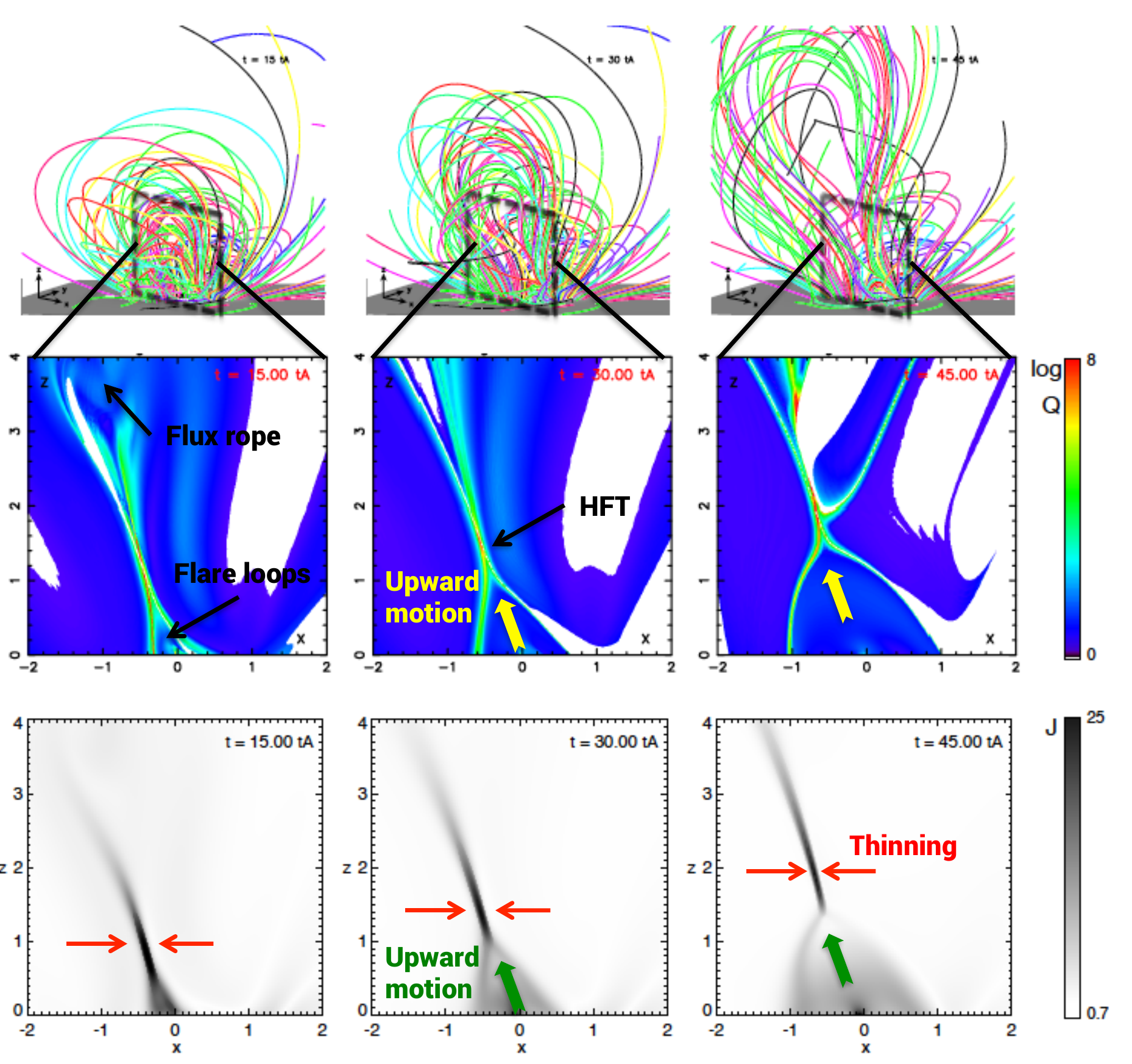}
\caption{3D representation and vertical cuts of an erupting flux rope. Results of the numerical simulation of a torus-unstable flux rope expansion with the OHM code \citep{Aulanier2012}. Top row: field lines showing the expanding magnetic field as time passes by (the times represented here are $t=15t_A, 30t_A,45t_A$). A 2D transverse cut (black dashed lines in the top row) of the QSLs is shown, for all three times, in the middle row. The QSLs delimitate different magnetic field domains related to the flux rope, flare loops and surrounding field. The region of the QSLs where the magnetic connectivity changes the most is indicated as the HFT (see \sect{QSLs}). A similar cut for the volumic current density $J$ is shown in the bottom row. The time evolution shows a thinning of the central current layer (indicated with red arrows), with an increased current density. The reconnection region and the top of the reconnected field lines move upward as time passes, as indicated with the yellow and green arrows \citep[adapted from][]{Janvier2013}.
}
\label{fig_9}
\end{figure}

\begin{figure}  %________________________ FIG ______________________________________    
\centering
\includegraphics[width=1\textwidth,clip]{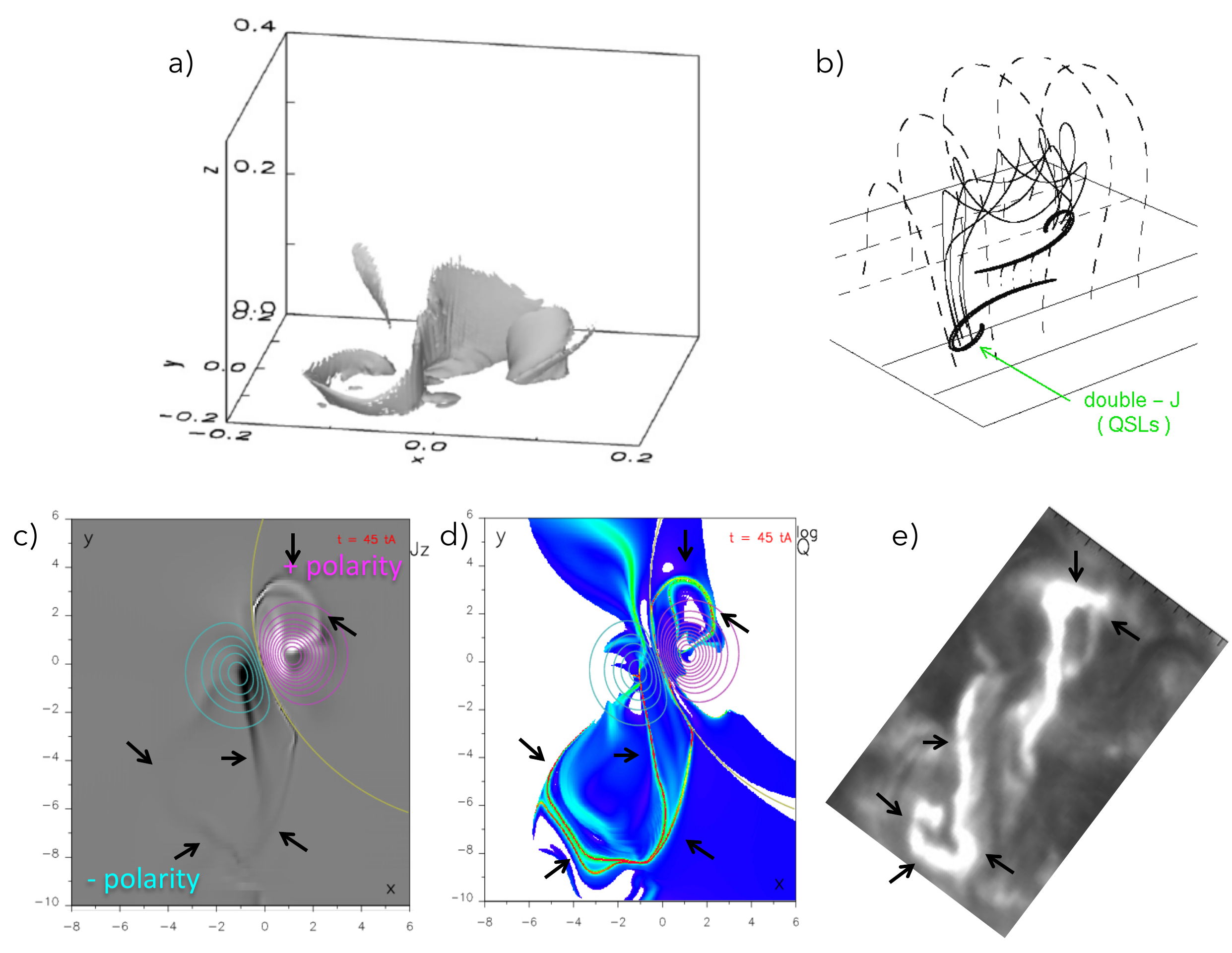}
\caption{Traces on the photospheric boundary of the QSLs and current density for an erupting flux rope. (a) The 3D volume of the current layer during a flux rope ejection, as simulated by \citet{Kliem2013} and similar to the simulation of \citet{Aulanier2012} and \citet{Janvier2013}. (b) A model of a flux rope (solid think line) underneath overlying arcades (dashed lines), showing the hooked, \Jshape-shaped QSLs (thick lines), as was first investigated in \citet{Demoulin1996}. (c) Top view of the photospheric ($z=0$) footprints of the vertical component of the current density vector $J_z$ in grayscale for the flux rope eruption simulation of \citet{Aulanier2012}. The magnetic polarities are shown in magenta (positive) and cyan (negative). (d) Same view for the photospheric footprints of the QSLs. The similar \Jshape-shapes for the current density and the QSLs are shown with the black arrows. (e) \Jshape-shaped flare ribbons during an eruptive flare \citep{Chandra2009}.
}
\label{fig_10}
\end{figure}

\subsection{Current layers associated with flux ropes}  %%%%%%%%%%%%%%%%%%%%%%%%%%%%%
\label{sect_currentlayereruptiveflares}

Global evolutions of current layers, in the context of solar flares, can be investigated with 3D simulations. {In the following, we focus on current layers during eruptive flares. In this context, the presence of QSLs is associated with the presence of an eruptive flux rope, as was discussed in \sect{QSLs}.} For example, in a series of papers, \citet{Aulanier2010,Aulanier2012} and \citet{Janvier2013} investigated the evolution of a torus-unstable flux rope and the underlying mechanisms of 3D reconnection. It was shown that the upward motion of the unstable flux rope away from the solar surface stretches the surrounding overlying arcades. This expansion motion is shown in \fig{9}, top row, where field lines are drawn at three different times during the evolution of the flux rope eruption (these lines have one fixed footpoint chosen at the beginning of the simulation). 

As it is quite hard to discern the different structures in the volume, the authors performed an analysis on both the QSLs and the current density structures within the simulated volume, which are shown respectively in the 2nd and 3rd rows of \fig{9}. Both the QSLs and the high current density regions are co-spatial. Throughout the flux rope evolution, the QSLs draw the frontiers between different connectivity region: the tear-drop shape structure (best seen in the left and central panels) indicates the volume associated with the flux rope (see also in \sect{QSLs}), while the $\Lambda$ region underneath marks the area where the flare loops are formed. Field lines filling the rest of the volume are overlying and surrounding arcades. 

The current layer is present along parts of the QSLs. This confirms previous analyses where current layers form where QSLs exist, but also that QSLs can exist without being necessarily related with current layers (\eg\ the tear-drop shape region is not seen in the current density transverse cut in \fig{9}). Interestingly, QSLs and current layers have a comparable evolution. First, their shapes are seen to evolve similarly: as time passes by, the inversed tear-drop shaped region moves upward, stretching the QSLs and the current region upward, while the $\Lambda$-shaped flare loop region grows (both are represented by the yellow and green thick arrows in \fig{9}). Furthermore, the current layer formed underneath the tear-drop structure thins (indicated with red arrows in \fig{9}, bottom row) while the current density and squashing degree $Q$ both grow in values. The location of the highest current density is nearly co-spatial with that of the highest values of $Q$ which define the so-called HFT \citep[although not exactly identical, as shown in Fig.2 of][]{Janvier2013}.

It is important to understand that in 3D, the current layer has a finite volume. The extent of this volume was shown in \citet{Kliem2013}, who performed an MHD evolution of an extrapolated unstable coronal flux rope. It is reproduced in \fig{10}a, where the volume  shows a similar structure as that found with the OHM simulation of \citet{Aulanier2012,Janvier2013}. \fig{10}a shows that the volume maps all the way down to the bottom boundary or photosphere, and reaches its highest height in the center of the volume (similar to the transverse cuts of \fig{9}, bottom row). The 3D volume is rather thin, and displays an \Sshape-shape. This shape is reminiscent of the double \Jshape\ found for the QSLs in the presence of flux ropes (\fig{10}b and \sect{QSLs}). In \fig{10}c and d, the photospheric $z$-component of the current density, $J_z$, is shown, along with the footprints of the QSLs. These are top views of the same structures shown in \fig{9}, middle and bottom row. Here, both the current density and the QSLs show the same structure, with the expected \Jshape-shapes, which are reminiscent of the flare ribbons typically seen during eruptive flares (\fig{10}e).

In such simulations, the current layer is not well resolved: this can be seen in \fig{9}, where the HFT region is much narrower than the current layer thickness. Indeed, the resistivity is quite large compared with solar values (to ensure numerical stability), so that the current diffuses much more than what would be expected in reality. Therefore, it remains difficult to properly address the physics of the current layer evolution, with respect to the development of instabilities such as the tearing mode. Recent investigations by \citet{Nishida2013} and \citet{Wyper2016} show studies of the behavior of the 3D current layer, with multiple shredding and plasmoid formation as would be expected from the 2D, well studied evolution. However, the complexity of the configuration in 3D \citep[see Fig.4 in][]{Nishida2013} shows that it is difficult to properly define the equivalent of magnetic islands seen in 2D. {This is because in 3D, plasmoids are not bounded by flux surfaces in the 2D system \citep[see \eg][]{Daughton2011}.} Still, this remains promising for further studies to understand the behavior of the current volume and link it to observational consequences of reconnection (see \sect{plasmoids}).

\subsection{Electric currents in observations, and their associations with flare ribbons and QSLs}  %%%%%%%%%%%%%%%%%%%%%%%%%%%%%
\label{sect_obscurrents}

%  {\S\bf Field changes during flares} \\
  
Already in the 1960s, changes in the magnetic field during flares were already reported by \citet{Severny1964b}. However, the temporal resolution at this time meant that each measurement was done far from the flare time, so that it remained difficult to conclude whether those changes were really flare-related or intrinsic to the evolution of the active region. 
Over the years, refinement in the spatial and temporal resolutions have brought more and more evidence of changes in the photospheric magnetic field during a flare. Some authors reported a decrease in the longitudinal magnetic field \citep[\eg][]{Kosovichev1999}, while some others \citep[see][and references therein]{Sudol2005,Petrie2010} reported both increase and decrease depending on the flaring regions investigated. Local changes in the horizontal photospheric magnetic fields were also recently reported \citep[\eg][]{Wang2002,Sun2012,Wang2012,Petrie2013}, while in those studies, the vertical field component does not change much.

%  {\S\bf Evolution of currents during flares} \\

Studies of electric currents during flares were already made with the first magnetograms. In particular, 
\citet{Moreton1968} already showed then that flare knots (defining the small areas of intense H$\alpha$ emission) arise near regions where strong electric current density were reported. 
Other authors have also reported extended patches of strong current densities with ground-based
magnetograms \citep[\eg][]{Hagyard1988, Hofmann1988, delaBeaujardiere1993}. Recently, \citet{Janvier2014} and \citet{Musset2015} investigated the evolution of strong electric current density regions with a higher spatial and temporal resolution given by HMI aboard the Solar Dynamics Observatory mission \citep[see the details of the instrument in][]{Schou2012}. 

\begin{figure}  %________________________ FIG ______________________________________    
\centering
\includegraphics[width=1\textwidth,clip]{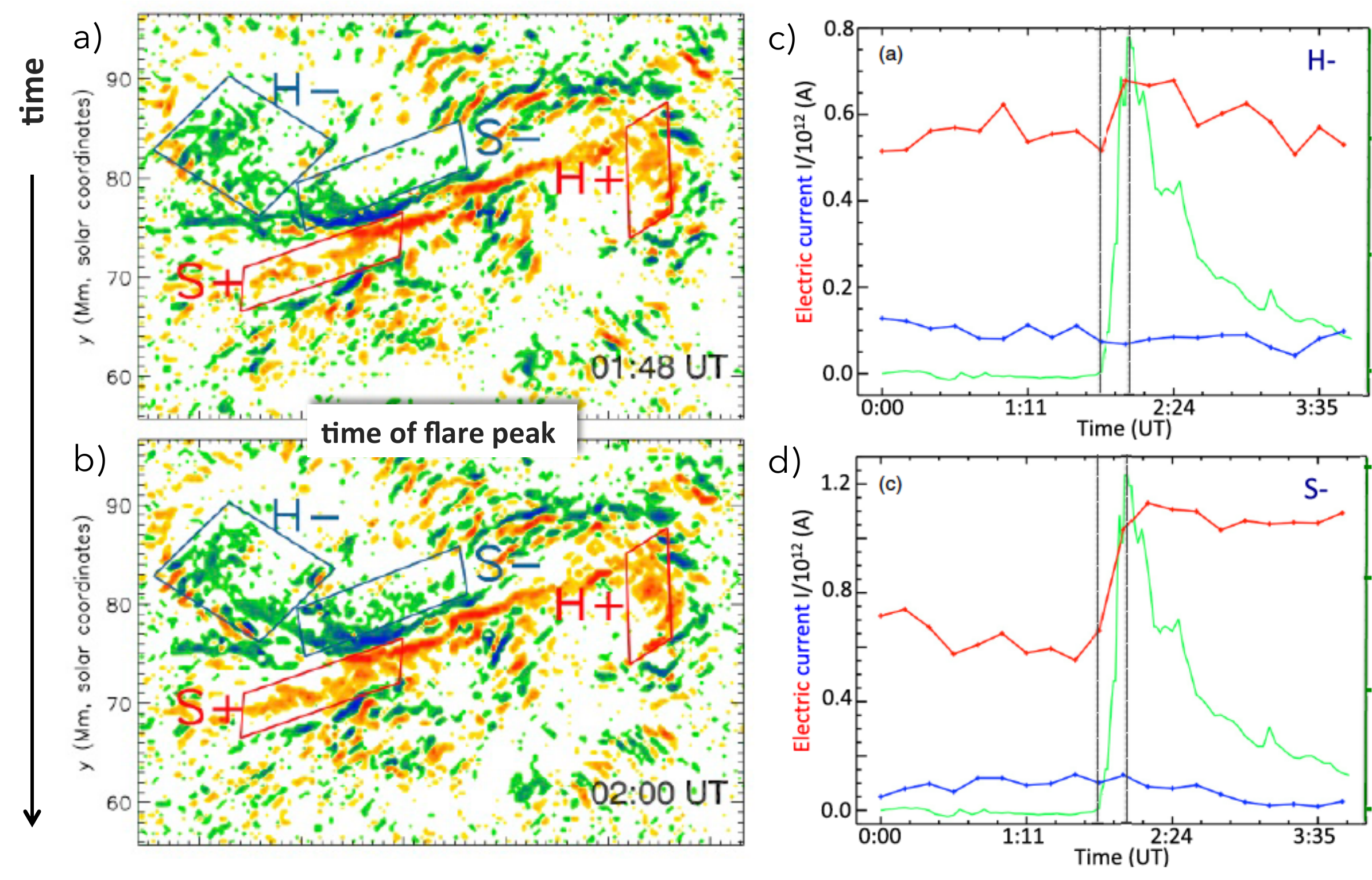}
\caption{Photospheric map (where the background noise has been removed) of the vertical ($z$) component of the current density at 01:48 UT (a) and 02:00 UT (b) on 15 February 2011 when an X-class flare was recorded. The time of the flare peak, from GOES Soft X-ray bands, is between the two snapshots shown here. The four squares are marked as areas where the strongest changes are seen before and after the flare impulsive phase. The regions marked with $S $ indicate the straigth part of the \Jshape-shaped current ribbon, while those marked with $H$ indicate the hook region of the \Jshape. (c-d) The light curve in the 335\AA\ filter of the AIA instrument aboard the Solar Dynamics Observatory \citep{Lemen2012} is shown in green, while the time evolution of the electric current $I$ (computed over the regions H- and S- defined in panels a and b) is shown in red (respectively blue) for the direct current (respectively return current, see text for details). The figure is adapted from \citet{Janvier2014}. } 
\label{fig_11}
\end{figure}

With the 12 min cadence measures of the X-class flare seen on 15 February 2011 in AR11158 (SOL2011-02-15T01:53), \citet{Janvier2014} used an inversion method of the observed polarized spectral profiles to obtain the three components of the magnetic field \citep[\eg][]{Bommier2007}. HMI data show that the photospheric traces of the vertical current component $J_z$ are similar to what is predicted from the shape of QSLs in the presence of flux ropes (see \fig{10}).
This shape is shown in \fig{11}a,b, where the $J_z$ component displays the expected thin and elongated \Jshape-shape. Interestingly, the current distributions are different, as they represent a time before (a) and after (b) the onset of the flare. In particular, in the regions highlighted with parallelograms (blue and red, marked H-,S-, H+, S+), one can see an increase in the current density values: the current ``ribbons'' are seen to elongate and spread, similar to the outward motion away from the inversion line of observed EUV ribbons. It is also possible to integrate the current density in the different highlighted areas in order to obtain the time evolution of the electric current $I=\iint I$d$x$d$y$. Its evolution is shown in the two areas H- and S-, where one can see that the sudden increase in the direct current is associated with that of the light curve, \ie\ with the impulsive phase (shown in red in \fig{11}c,d). A direct current is defined as parallel/anti-parallel to $\vec{B}$ for a positive/negative current helicity. The increase in the current density remains rather stable during the decreasing phase of the flare.

This result may at first sound counter-intuitive, since we would expect that the current density decreases, as a flare leads to a reconfiguration of the magnetic field where the free energy has decreased, \ie\ closer to a potential state. However, there is a competition between two mechanisms: the ideal instability which triggers the field evolution, and the subsequent reconnection. The first mechanism implies the increase of the current magnitude in the current layer, a generalisation of what is known in 2D \citep{Lin2000}. Then, an increase of current magnitude is expected when reconnection cannot dissipate fast enough the accumulated current density in the current layer. With recent observations, as shown in \fig{11}, we are now able to witness the collapse of the current layer and the implied increase in electric current, which corresponds to the onset of the flaring phase (as fast reconnection is then triggered).

%  {\S\bf Current ribbons in more complex flaring regions} \\

\begin{figure}  %________________________ FIG ______________________________________    
\centering
\includegraphics[width=1\textwidth,clip]{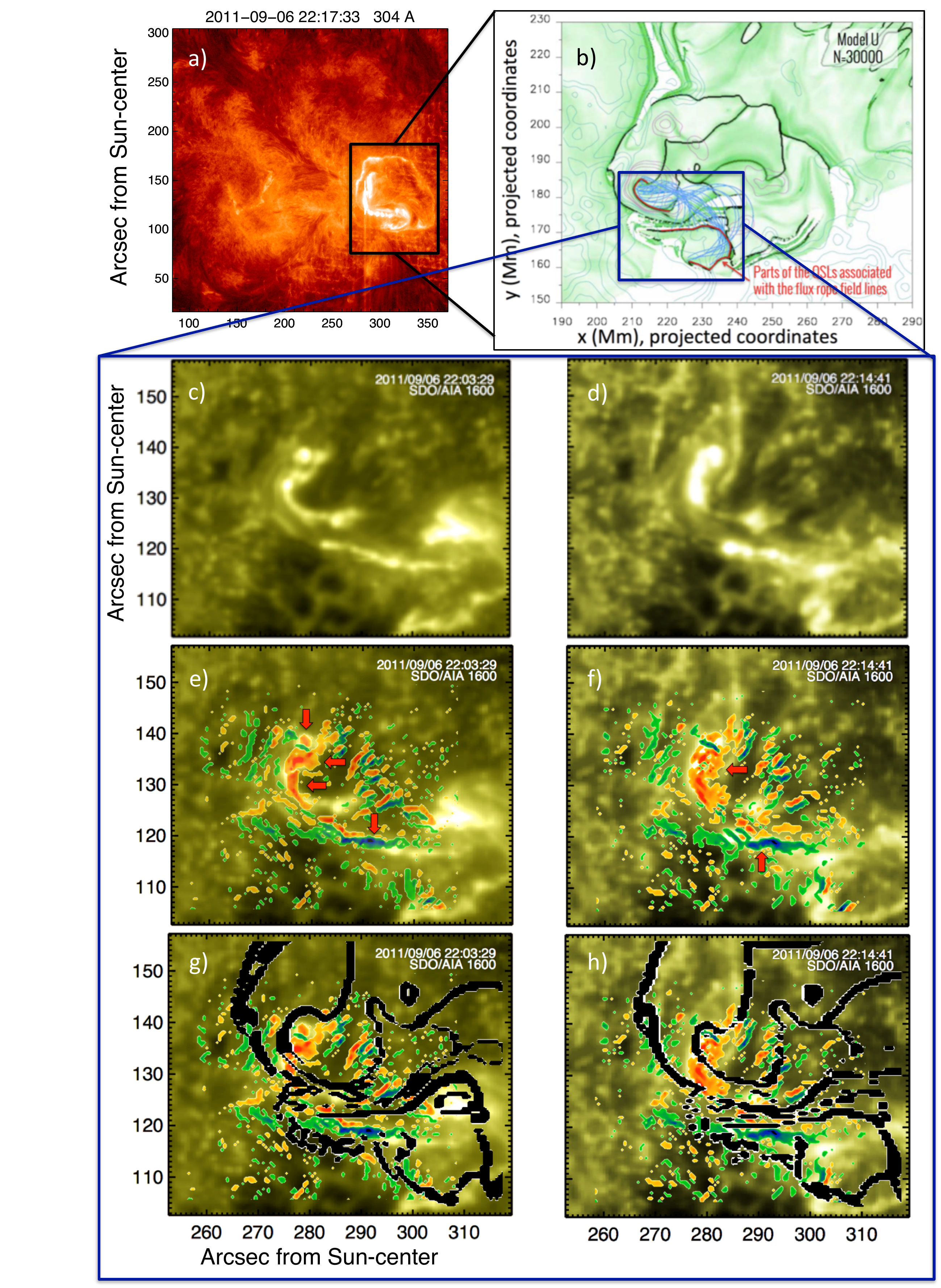}
\caption{Current ribbons and QSL comparison in a complex flaring region. (a) Overview of the X flare region of 6 September 2011 in the 304\AA\ channel of AIA aboard SDO. A large-scale circling flare ribbon (rectangle box) indicates the presence of a fan-like structure, while the most southern ribbon displays a hook shape typical of flux rope ribbons. (b) QSL photospheric map of the zoomed region (black box in panel a) showing similar structures as the flare ribbons. In particular, a flux rope found in the extrapolated magnetic field (blue lines) is anchored in QSL regions displaying the typical \Jshape-shapes on both sides of the inversion line. A zoom on the two flare ribbons associated with the flux ropes are shown in the 1600\AA\ filter before (c) and after (d) the impulsive phase. (e-f) An overlay of the same region with the current density obtained with the HMI data is shown for the same times. (g-h) Same overlays adding the local QSLs from the extrapolation, showing a good agreement in the shape and the location of the QSLs (extrapolation), currents (HMI) and EUV flare ribbons. Adapted from \citet{Janvier2016}.  } 
\label{fig_11b}
\end{figure}

Following up on this study, \citet{Janvier2016} investigated the more complex flaring region of 6 September 2011 (SOL2011-09-06T22:20). This region, as shown in \fig{11b}, shows a complex network of flare ribbons compared with a typical 2-ribbon eruption.
A careful investigation with a magnetic field reconstruction model showed that a flux rope was present, embedded in a topology {reminiscent of a null point configuration, although a stable null point was not found in the extrapolations}.
A parasitic polarity is responsible for the spine/fan like configuration, similar to what is normally found in the presence of magnetic null points \citep[see][for an example of flare ribbons in the presence of a magnetic null point]{Masson2009}. The connectivity mapping of the domain was provided by the technique developed in \citet{Pariat2012} and a novel technique for QSL calculations in 3D as described in \citet{Tassev2016}. As such, when the flare started, the flux rope ejection was accompanied with reconnection occuring at the quasi-spine/fan configuration, with some flare ribbons tracing the photospheric traces of the fan dome. Still, the two opposite ribbons associated with the erupting flux rope displayed the typical \Jshape-shape, showing that even in complex configurations, flux rope eruptions display similar outcomes as expected from analytical/numerical models of twisted configurations.

Then, the QSLs computed from the magnetic field were compared with the locations of the flare ribbons seen in different filters, as well as the locations of the electric current density regions discussed above. It was then found that the QSLs associated with the flux rope location \citep[as found with the flux rope insertion method, described in][]{VanBallegooijen2004} displayed a similar \Jshape-shape feature as those reported in analytical and numerical studies. 

Following this analysis, the magnetic model of the flaring active region was forced to evolve via a magnetofrictional code. This technique has been applied to a series of active regions, as reported in \citet{Savcheva2016}. In the case of the 6 September 2011 event, it was shown that the QSLs spread away from the inversion line, for magnetic models that included an unstable flux rope. Moreover, there were local changes similar to that found in the current density regions (investigated from HMI data during the same event) and the flare ribbons seen in the different filters of AIA.

As such, with different methods (magnetic field reconstruction and evolution, HMI and AIA), the authors were able to find the same consistent morphologies and evolution in flare ribbons, QSLs and electric current densities, as predicted in analytical (morphology) and numerical models (morphology and evolution). 
It is quite interesting that the standard flare model in 3D, derived from MHD simulations, is able to reproduce quite well the localisation of the impact of energetic particles as seen as flare ribbons, although it actually does not solve the transport of energy to the lower layers of the Sun's atmosphere. Then, even though one can argue that MHD models do not take into account this transport (\eg\ with energetic particles or waves), the fact that the evolution of the large-scale field is able to explain the observed features means that the MHD assumption is a strong and valid assumption to explain the large scale features of the Sun's atmosphere.

%%%%%%%%%%%%%%%%%%%%%%%%%%%%%%%%%%%%%%%%%%%%%%%%%%%%%%%%%%%%%%%%%%%%%%%%%%%%%%%%%%%%%
\section{The consequences of 3D reconnection in solar flares} %%%%%%%%%%%%%%%%%%%%%%%%%%%%%
\label{sect_consequences}

\subsection{Slipping reconnection}  %%%%%%%%%%%%%%%%%%%%%%%%%%%%%
\label{sect_slipping}

%  {\S\bf What is slipping reconnection?}\\

\begin{figure}  %________________________ FIG ______________________________________    
\centering
\includegraphics[width=1\textwidth,clip]{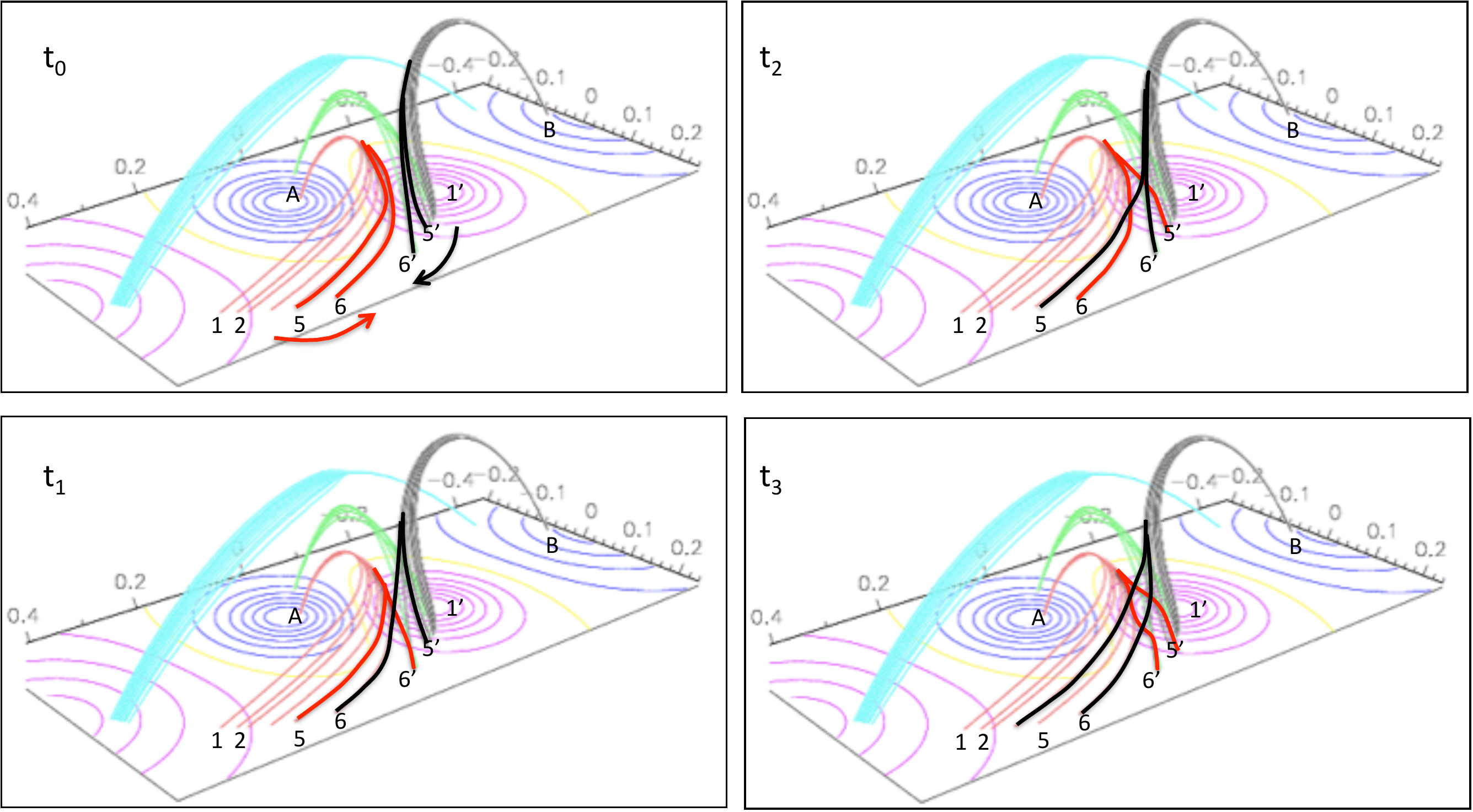}
\caption{Representation of slipping field lines at different times in a numerical simulation. Four sets of field lines are represented, all defined from the negative polarity. The neighbouring anchorpoints of the cyan and black (respectively red and green) field lines in the same polarity, and the diverging locations of the corresponding footpoints in the positive polarity show that the connectivity remains continuous, while strongly diverging. The four thick red and black lines are defined as departing from local area A (red lines A5 and A6) and B (black lines B5' and B6'). At four different times, we look at the changes in the connectivity while those field lines are reconnecting with each other. The continuous change of connectivity gives an apparent slipping motion, indicated with the coloured arrows at $t_0$. \citep[adapted from][{see online for supplementary material showing the slipping motion in an animated gif.}]{Aulanier2006}} 
\label{fig_12}
\end{figure}

%In \citet{Syrovatskii1966}, it was already suggested that in regions where an electric induction field appears,  \ie\ where the freezing-in assumption no longer stands, lines of force are said to move, slipping with respect to the medium. 
Reconnection without a null point but in the presence of a parallel electric field due to the stress of magnetic fields near regions of strong distortions of the magnetic connectivities, was generalised in a series of papers and applied to the understanding of solar flares (see \sect{QSLs}). In such a case, magnetic field lines are also seen to flip \citep{Priest1992} or slip \citep{Priest1995}, as they undergo a continuous change of connectivity in the reconnection layer.

To illustrate the notion of slipping field lines, let us take a set of field lines in \fig{12} leaving the photospheric surface at similar locations: the red field lines leave from a footpoint neighborhood A and the black lines a footpoint neighborhood B. {In the following, we detail a connectivity change between two field lines at each time step, however note that this is only for a description convenience, since all field lines passing through the current volume would undergo continuous reconnection \cite[see \eg][]{Priest2003}}.  Note that the configuration is similar to \fig{6}, where the photospheric traces of the QSLs are shown in \fig{6}b. We pick two field lines represented in thick lines, red A5 and A6 and black B5' and B6' (where the digits indicate the other footpoints). Note that the connectivity between the two positive magnetic polarities (represented in magenta) is continuous, although drastically changing as can be seen by the jump between 6 and 6'. Then, at the onset of reconnection, line A6 reconnects with its neighbouring field B6': at time t1, we can then define lines A6' and B6. As reconnection continues, we then have at $t_2$ newly formed A5' and A6, as well as B5 and B6'. Reconnecting field lines then give an apparent sliding motion: the field line defined at $t_0$ as A5 has become A6 at $t_2$, then A6' at $t_3$, so that its anchoring footpoint in the positive polarity has an apparent motion going from points 5, 6 then 6'. 
It is the successive reconnections that is named the ``slipping'' apparent motion of field lines.\\
  
\begin{figure}  %________________________ FIG ______________________________________    
\centering
\includegraphics[width=1\textwidth,clip]{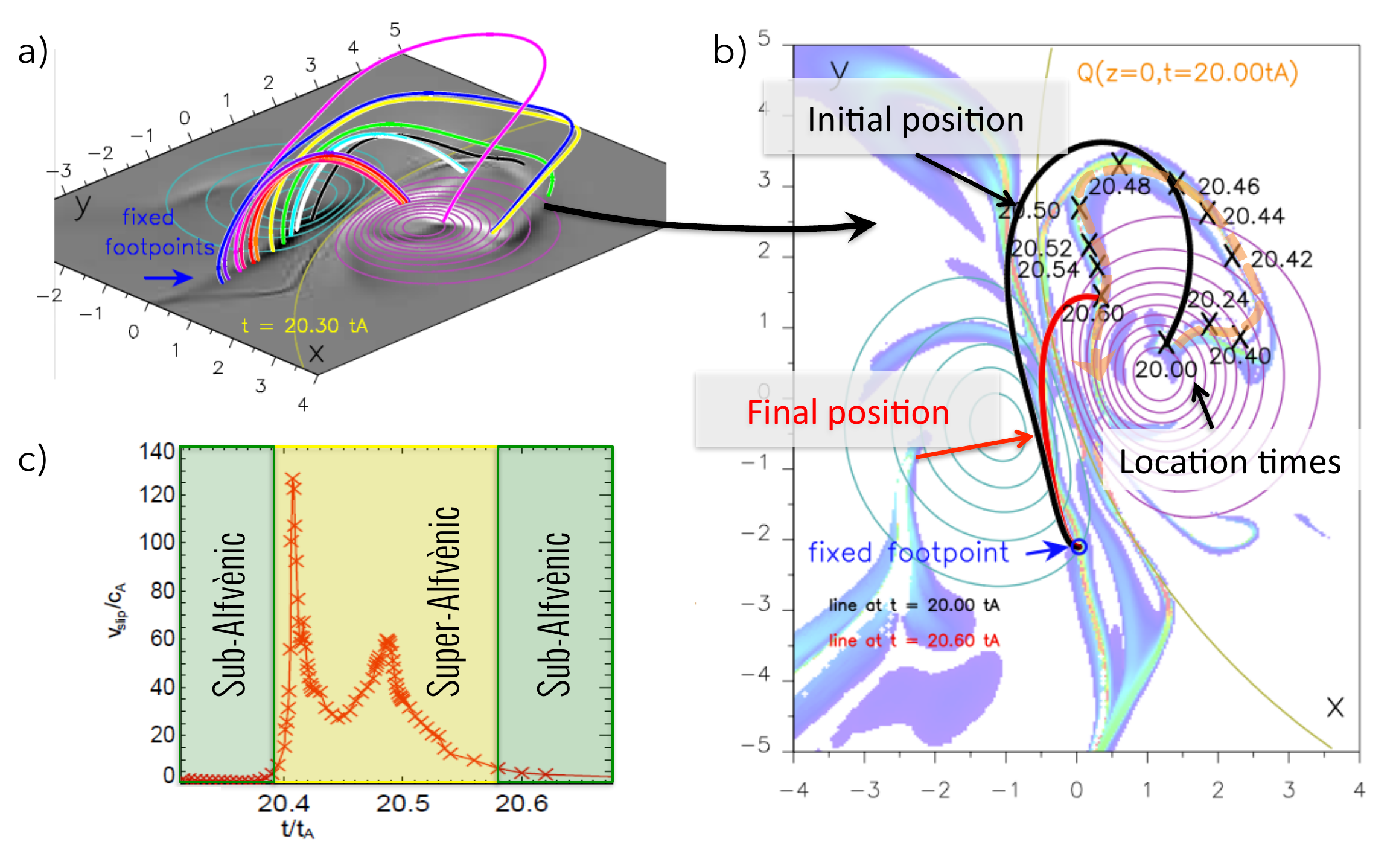}
\caption{Calculation of the slipping speed. (a) Set of slipping field lines at a given time from an eruptive flare simulation \citep[see][]{Janvier2013}, which anchoring point is indicated as fixed in the negative polarity. (b) The footpoint locations of the moving yellow field line of panel (a) at different times in the simulation are indicated by crosses, on an overlay of the QSL photospheric map. The initial position of the yellow field line is indicated as a black line, while its final position (when reconnection ends) is indicated as a red line (different color coding as panel (a)). The path taken by the moving footpoint is indicated with an orange dashed line. (c) Time evolution of the local speed of the moving footpoint (normalised by the Alfv\'en speed). Two regions are indicated. In green, we find the times when the motion is sub-Alfv\'enic ($v_\mathrm{slip} \le c_A$), \ie\ at the beginning and at the end of the QSL crossing. In yellow, we find the times when the motion is super-Alfv\'enic ($v_\mathrm{slip} \ge c_A$, slip-running motion, \ie\ in the core region of the QSL where the connectivity gradient is the highest).
} 
\label{fig_13}
\end{figure}

At any time during the evolution of the magnetic field, one can define a specific field line by fixing one footpoint at the photospheric surface. More generally, in a numerical simulation where the bottom boundary evolves, one can define a field line by following its footpoint motion. That is the case for the yellow line shown in \fig{13}a. We fix one of its footpoint in the negative polarity and at any timestep in the simulation, field integration gives the location of the corresponding footpoint in the positive polarity.

If this field line is reconnecting, its change of connectivity can be followed with time: the corresponding footpoint location can be reported on a map such as in \fig{13}b. Note that the field line is not jumping from one photospheric location to another: rather, the footpoint is continuously changing of locations because of the continuous change of connectivity (as there is no separatrices). Knowing the distances between two connectivity changes and the time step, one can then define a local slipping speed $v_\mathrm{slip}$: its variations during the reconnecting time interval are shown in \fig{13}c. 

Interestingly, the slipping speed (or connectivity change rate) is evolving in time. When compared with the Alfv\'en speed, we remark that the ratio can either be less than 1 or much larger: in \citet{Aulanier2006}, the authors defined a sub-Alfv\'enic slipping motion as a motion for which the slipping speed is less than the Alfv\'en speed, and inversely for a super-Alfv\'enic motion. In the latter case, the slipping motion is referred to as a slip-running motion. In the example of \fig{13}, the chosen field line slipping speed evolves from 0 to a sub-Alv\'enic speed, then a super-Alv\'enic speed. This actually represents the crossing of a QSL: further away from the QSL (and the current layer), the field line is not reconnecting ($v_\mathrm{slip}=0$). As the field line crosses the QSL, it undergoes reconnection with its neighbouring field lines (such as in \fig{12}). As the connectivity gradient increases while passing through the QSL, the field line reconnects with a neighbouring one which footpoint is even further away. This is seen as a larger ``jump'' in the newly reconnected field line footpoint location, and a greater value of its slipping speed. In the core of the QSL, where the gradient is the greatest, the slipping speed is the largest, corresponding to a slip-running reconnection regime. As the field line is now exiting the QSL region, its slipping speed drops to sub-Alfv\'enic values, until reaching 0 again.

%  {\S\bf Link with the norm}\\

What dictates the slipping motion of field lines? The connectivity gradient is an important factor (and as such, is described by the geometry of the field, or QSLs), but so is the dissipation process within the current layer: as explained above, when passing through the dissipation volume, magnetic field lines reconnect with a nearby field. Successive reconnections are then dictated by two aspects: the physical dissipation term which appears in Ohm's law (\eg\ the resistivity) and the geometry, which provides an information on the connectivity of the neighbouring field with which field lines are going to reconnect with. This latter aspect was studied in great details by \citet{Janvier2013}, who found that the field line mapping norm (\eq{mappingnorm}) is essential in the speed of the slipping motion. Indeed, since slipping field lines are defined from one of the two footpoints, their slipping apparent motion will depend on the connectivity of their neighbouring field lines, which is given by the mapping norm. In other terms, the time evolution of the slipping speed, shown in \fig{13}c, has the same profile as the spatial distribution of the mapping norm $N$:
\BE
v_{\mathrm{slip}}=\alpha N
\EE
where $v_{\mathrm{slip}}$ represents the {apparent} speed of the moving footpoint and $N$ the mapping norm. The parameter $\alpha$ is itself a coefficient related with the reconnection rate within the current layer. 
Note that again, reconnection in the presence of QSLs generalises the results found in 2D: when the squashing degree $Q$ becomes infinite, in which cases QSLs generate real separatrices (in the topological sense), the slipping speed becomes infinite (the field line footpoint only ``moves'' from an initial point to a final point) as $N=\infty$.\\

%  {\S\bf Slipping as seen in observations.}\\

\begin{figure}  %________________________ FIG ______________________________________    
\centering
\includegraphics[width=1\textwidth,clip]{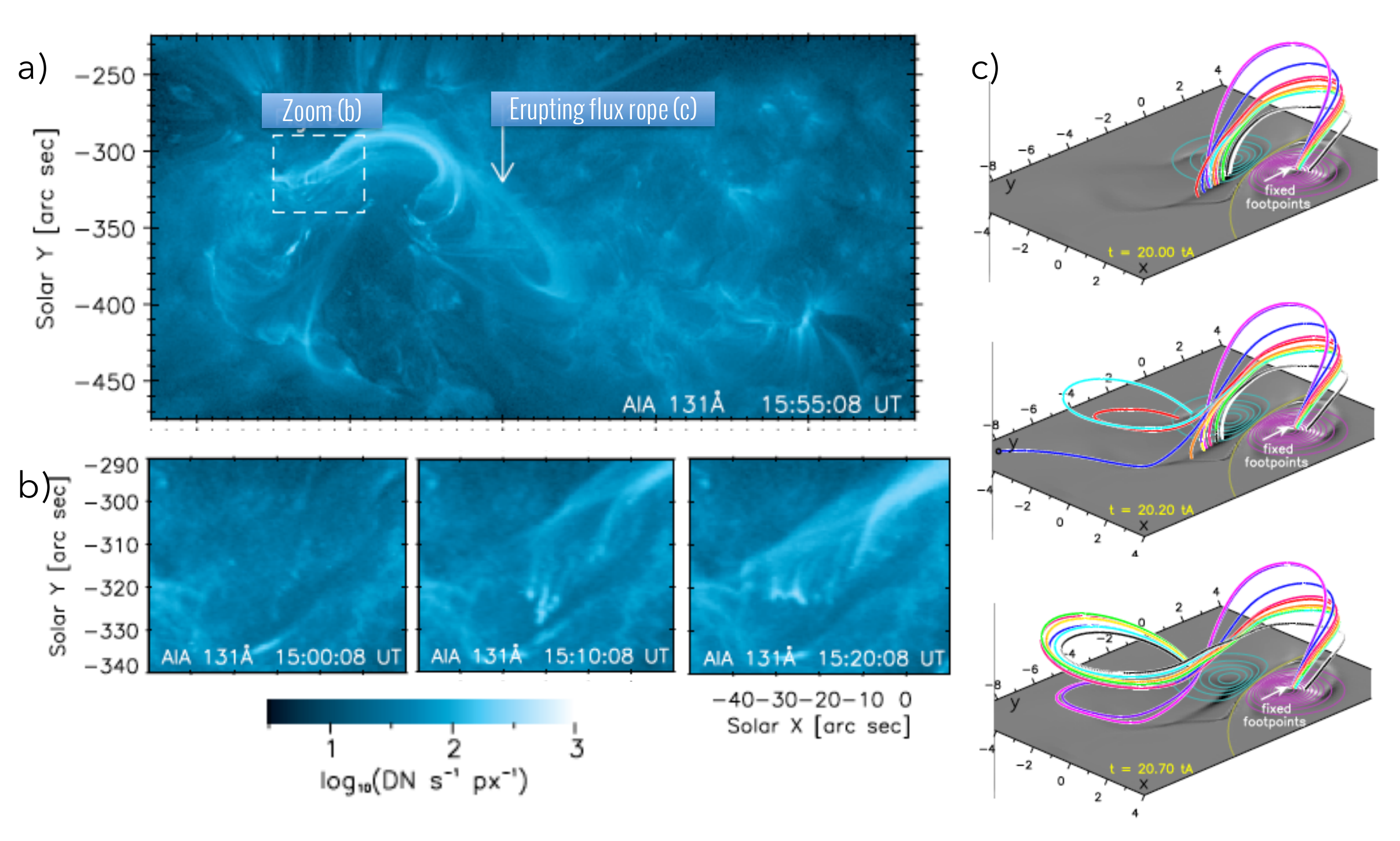}
\caption{Evidence of apparent slipping motion during the X-class eruptive flare of 12 July 2012 (SOL2012-07-12T16:49). (a) Overview of the region where hot loops are seen in the 131\AA\ (Fe XIII and Fe XXI), including an eruptive set of expanding loops \citep[see Fig.5 in][]{Dudik2014}. The zoom region is shown in (b) at earlier times, where coherent, unidirectional motions of kernel brightenings and apparent coronal loop motions are seen. (c) The analogy can be made with the slipping motion of magnetic field lines from a numerical simulation of an eruptive flares, where at different times reconnecting flux rope field lines are seen to slip. \citep[adapted from][]{Dudik2014}.
} 
\label{fig_14}
\end{figure}

So far, we have only discussed slipping reconnection in terms of magnetic connectivity changes. However, these changes have consequences on the surrounding plasma. In the dissipation layer, the magnetic field energy is converted in other forms of energy such as kinetic energy of particles (later seen as Hard X-ray signatures) and heat. The apparition of bright kernels and flare ribbons at chromospheric altitudes during flares are understood as a transport of energy from the reconnection site in the corona to the chromosphere \citep[see \eg][]{Graham2015}. Whether this energy is transported along field lines under the form of high energy particles thermal front or waves is still debated \citep{Kerr2016}.
Since in the presence of QSLs, and as seen above, the change of connectivity is continuous and leads to a slippage of field lines, the apparently moving magnetic footpoint shown in \fig{13} may have consequences on the sequential timing of the disturbances seen at the chromosphere. 

With the high temporal cadence instrument AIA aboard Solar Dynamics Observatory, \citet{Dudik2014} were able to look in details at the motion of kernels seen during an eruptive flare. For the first time, a detailed analysis of these kernels at the highest time cadence of the instrument (12s) showed that their motions are coherent: the kernels are seen to lit up successively in a privileged direction. They are seen in different parts of the two ribbons appearing during the eruptive flare and accompanied by an apparent slipping motion of coronal loops \citep[which was already pointed out in][with X-ray observations of Hinode]{Aulanier2007}. 

It may seem surprising that similar consequences are seen in both numerical simulations and observations. However, let us not forget that the plasma in the photosphere is at least $10^9$ times denser than the corona: disturbances from the corona to the photosphere would have great difficulties to affect and move the extremely massive plasma of the photosphere. As such, the field lines that would map the magnetic field from the photosphere can be said to be ``lined-tied" to the photospheric surface: hence, a similar interpretation as the simulation can be made.

Recently, several papers have reported similar analyses during eruptive flares showing the motion of kernels \citep{Li2014,Zheng2016,Li2016,Sobotka2016}. This should not be confused with the ``zipper''-propagation seen during prominence eruptions such as in \citet{Tripathi2006}: there, the locations of brightening kernels are seen to be related with the eruption direction of the large magnetic structure. Depending on whether the prominence lifts off symmetrically or asymmetrically, the flare kernels/ribbons appear sequentially at different locations due to the propagation of the reconnection site. With the intrinsic slipping motion of coronal loops, we can therefore expect to see kernel brightenings appearing at the footpoints of both the erupting flux rope and the flare loops. This was reported in \citet{Dudik2016}, where the authors showed observations of moving kernels belonging to both flare loops and the expanding flux rope.

Although kernels are related with high energy particles that can travel much faster than the bulk plasma, the plasma in the corona can also respond to the successive change of connectivity. Whether it is heating directly from the dissipation layer, or so-called evaporation from the chromosphere, newly formed coronal loops can be filled with hot plasma that is emitting in EUV and X-rays. However, the fastest speed at which the information of the change of connectivity can travel, in the plasma, is bound by the Alfv\'en speed: then, the apparent motion of field lines can be seen in observations when the slippage is sub-Alfv\'enic.

\subsection{Flare loops and flux rope}  %%%%%%%%%%%%%%%%%%%%%%%%%%%%%
\label{sect_flareloops}

\begin{figure}  %________________________ FIG ______________________________________    
\centering
\includegraphics[width=1\textwidth,clip]{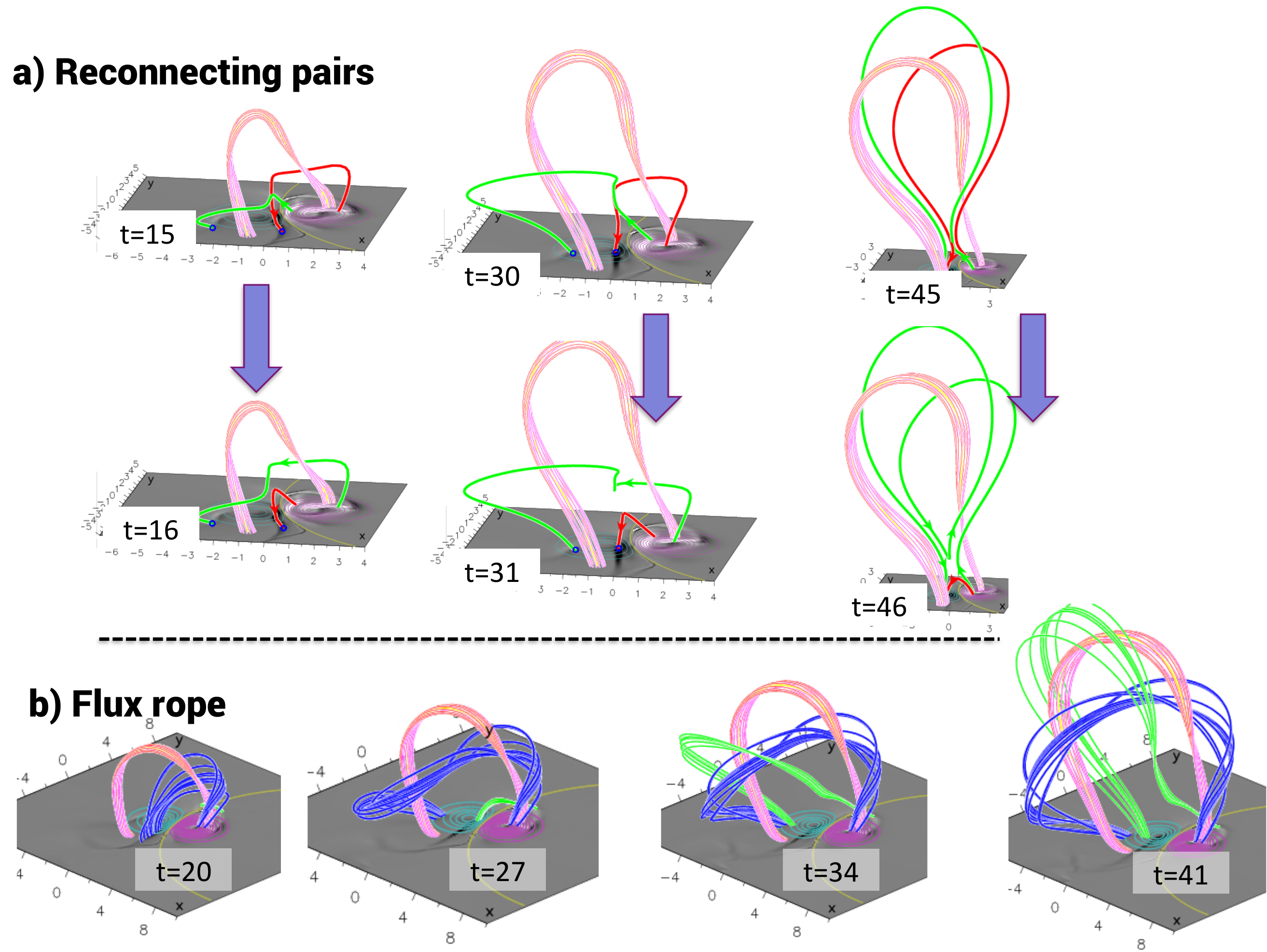}
\caption{Consequences of reconnecting pairs of field lines. (a, first row) Two pairs of field lines (red and green) chosen at different times and reconnecting with each other. (a, second row) One Alfv\'en time later, newly reconnected field lines are the flare loop (in red) and a green field line that surrounds the flux rope (its core is represented in pink). As time goes by, the flux rope field line forming on the outside becomes more twisted (\eg\ at times 45-46 $t_{A}$) and the flare loop is less sheared. (b) Time evolution of selected neighbouring field lines in blue and green that undergo reconnection. The blue field lines reconnect earlier than the green ones, and both form the successive layers of the flux rope envelope. Adapted from \citet{Aulanier2012}.} 
\label{fig_15}
\end{figure}

%{\S\bf Flare loops} \\
Coronal loops in the Sun's atmosphere provide direct evidence of the consequences of magnetic reconnection during flares. They can be seen as strongly emitting features within the soft X-ray and extreme-ultraviolet (EUV) ranges. They typically appear after the onset of the flare, and can be seen throughout the impulsive phase and the decaying phase of the flare. These flare loops are seen in both confined and eruptive flares \citep{Svestka1986}. In the latter case, flare loops typically form at higher altitude as time passes by, as the energy release site moves upward \citep{Liu2004,Warren2011}. 
Time sequences of flares also indicate that flare loops formed at the beginning of the flare have a more pronounced shear, evidenced by comparing their direction with that of the local polarity inversion line (PIL), than those formed later \citep{Asai2003,Su2007}. In \cite{Aulanier2012}, the authors investigated the evolution of the shear from strong values to weak ones by comparing observations with a numerical model, and showed that coronal loops reconnecting later in time were also closer to potential loops, due to their initial low shear as well as to the dynamics of the ejected flux rope that decreases the shear of reconnecting field lines (see \fig{15} which illustrates this dynamics). 
The flare ribbons lie at the feet of those dense loops. We often refer to an outward motion of the flare ribbons, away from the polarity inversion line, however, this motion is only apparent as it is related with the progressive lighting up of the regions where newly reconnected flare loops are anchored, as reconnection propagates.\\

%{\S\bf Flux rope} \\

In eruptive flaring active regions, evidence of an already present twisted structure can be found. This is seen as the presence of a hot bundle of $S$- or $J$- shaped loops referred to as a sigmoid, which are generally emitting in hard X-rays \citep{Green2007,Green2009, Gibson2006,Tripathi2009}. In a series of paper, \cite{Savcheva2009} and \cite{Savcheva2012,Savcheva2012b,Savcheva2014} showed that sigmoids are recurrent in flaring active regions. This was also investigated by \citet{Nindos2015}, by using different AIA channels (131, 171 and 304 \AA) who concluded that flux ropes and flux-rope like structures can be seen prior to eruptions in 30 to 50 percent of the cases. Such studies clarify the existence of flux ropes before the eruption, although direct observations are difficult, and hence  the presence of a flux rope cannot be verified directly until accessing quantitative measurements of magnetic fields in the corona.
The existence of flux ropes prior to flares is corroborated with findings of numerical simulations. Photospheric motions such as diffusion and flux cancellation permet creation of erupting flux ropes, as was modelled in 3D simulations \citep{Amari2003,Aulanier2010}. The study of the destabilization of flux ropes, at the origin of eruptive flares, is out of scope of the present paper, and we refer the reader to the review of \citet{Schmieder2013}.

Upon the eruption, these structures can be observed in coronagraphic observations as a three-part structure \citep[\eg][]{Cremades2004} but also more recently in low coronal observations. They are observed off limb as large structures with a cavity \citep{Gibson2010,Aparna2016}. Thanks to numerical simulations, it is possible to probe into the evolution of flux ropes. As the magnetic field of the active region undergoes reconnection, creating flare loops, the counterpart of this reconnection process is the formation of the flux rope envelope: the reconnected field forms field lines that are wrapping around the core of the flux rope. This is shown in \fig{15}, where pairs of field lines are shown to end up on the one hand as flare loops and on the other hand as part of the flux rope envelope. Those field lines then become part of the flux rope, and the latter grows in size (\fig{15}b). As such, the flux rope is always in constant magnetic flux evolution during the flare.
This evolution can also be seen by studying the dimming regions, generally associated with the footpoint of the flux rope \citep[see][and references therein]{Webb2012}. In particular, the evolution of these regions gives a good indication of the flux content in flux ropes \citep{Qiu2007}. In some cases, those dimming regions are also seen to have a strong-to-weak shear evolution similarly to the flare loops, as is also shown in simulations such as in \fig{15} \citep{Miklenic2011}.
 Also, the flux rope that is seen to evolve in the Sun's corona can be linked with the interplanetary medium, where direct in situ observations give us a quantification of the flux and the rotation of the magnetic field, that can then be used to find the properties, such as the morphology of the flux rope, its flux and twist content \citep[][]{Dasso2005, Mandrini2005, Hu2014}.

\subsection{Plasmoids and outflows}  %%%%%%%%%%%%%%%%%%%%%%%%%%%%%
\label{sect_plasmoids}

%  {\S\bf Plasmoids in numerical simulations}\\
   
\begin{figure}  %________________________ FIG ______________________________________    
\centering
\includegraphics[width=0.9\textwidth,clip]{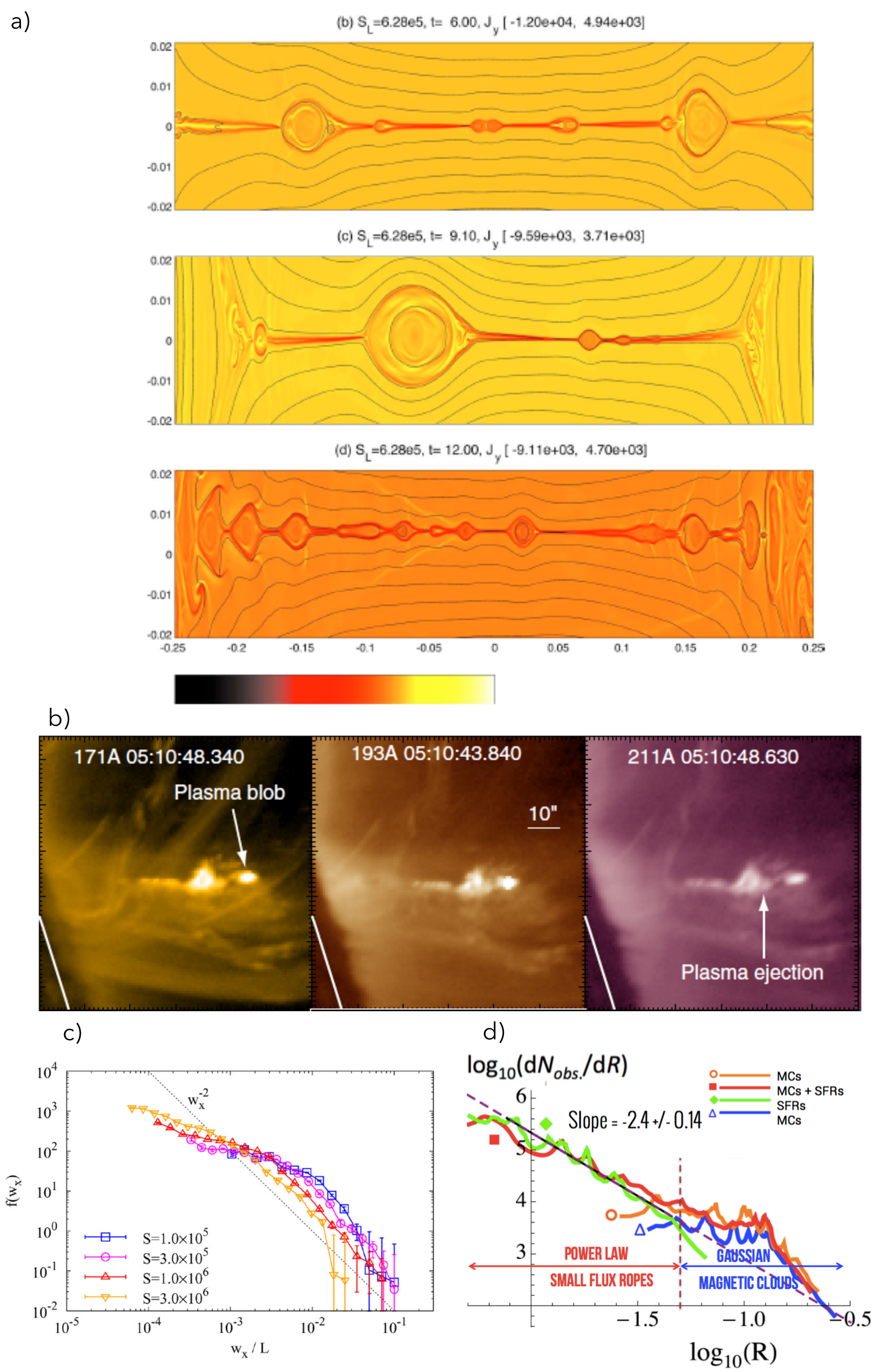}
\caption{Plasmoids in numerical simulations and observations. (a) Numerical simulations of \citet{Bhattacharjee2009}, where the formation and coalescence of plasmoids can be seen, creating large magnetic islands. (b) Remote-sensing observations of the Sun's corona during an eruption, where plasma blobs are seen during the rising phase of a flare and above the flare loop top \citep[adapted from][]{Takasao2012}. (c) Power law found in the distribution of the size of plasmoids in computer simulations, from \citet{Loureiro2012}. (d) A similar power law is found for small flux ropes directly observed in the interplanetary medium, from the study of \citet{Janvier2014b}. }
\label{fig_16}
\end{figure}

Numerical simulations of magnetic reconnection have shown that the evolution of unstable current layers (such as the tearing mode) leads to the formation of magnetic islands in 2D or plasmoids, that can be understood as small flux ropes in 3D. 
They are represented in 2D by nested magnetic field lines (see \fig{16}), while in 3D it is much more difficult to define what a plasmoid actually is. Indeed, in 3D, what would look like a coherent entanglement of field lines in one location may not be actually so different from the surrounding magnetic field in other locations in the 3D volume, {as discussed in \sect{currentlayereruptiveflares}} \citep[see for example the magnetic field rendering of the 3D simulation in][]{Daughton2011, Nishida2013}. {Recent numerical simulations such as by \citet{Baalrud2012,Wyper2014} provide invaluable tools to comprehend the link between 2D and 3D plasmoid dynamics.}

The formation and the evolution of plasmoids were invoked as an essential component in the impulsive, fast reconnection regime encountered in the solar corona: as a plasmoid is formed, the current layers become thinner, leading to a new Sweet-Parker regime that can become unstable to a secondary instability \citep{Shibata2001}.
As explained in \cite{Edmondson2010} and \citet{Lynch2016}, the plasmoid formation is a robust and universal process found in varied plasma simulations. It ranges from MHD to more detailed simulations taking into account particle effects. As they explain, inflows and outflows are determined by the global geometry of the system (which dictates the geometry of the current sheet, \ie\ the region where the frozen-in condition breaks down) and the resistivity/particle effect modelled (which dictates the diffusion scale). Since the system is also determined with other constraints such as conservation of mass and magnetic flux, the creation of magnetic islands and the further tearing of current sheets, such as in the plasmoid instability scenario, allows the system to introduce new scales to accomodate both the global constraints and conservation laws.

This subsequent nonlinear evolution from a tearing unstable current layer has since then been investigated in many different simulations. They indeed show the nonlinear evolution of unstable current layers into a Sweet-Parker regime and a further secondary plasmoids creation regime \citep[\eg][]{Loureiro2005,Samtaney2009, Ni2010, Militello2014}. This was also confirmed in laboratory experiments by \citet{Liang2007}.  Because plasmoids can grow, either from the subsequent evolution of the instability (continued reconnected flux accumulation) or by coalescence of magnetic islands \citep{Hayashi1981}, large plasmoids can be generated. Then, it is possible to study statistically the occurence of plasmoids from different sizes (see \eg\ the simulation of \citet{Bhattacharjee2009} in \fig{16}a). In particular, \citet{Fermo2010}, \citet{Uzdensky2010},  and \citet{Loureiro2012} discuss the formation of large plasmoids (also coined ``monster'' plasmoids by the two latter), which were also found by \citet{Lynch2016} in simulations of current sheets during eruptive flares. Such simulations are very interesting for comparisons with direct or indirect observations of plasmoids, both in the Sun's atmosphere and the heliosphere.\\

  %{\S\bf Plasmoids in remote-sensing observations}\\
  
The main consequences of magnetic reconnection are directly seen in remote-sensing observations as a large change in bolometric measurements (\ie\ light curves), ranging from microwaves to X-rays and even $\gamma$-rays. Then restructurations of the magnetic field as seen above, as well as outflows of structures such as plasmoids, have also been reported. For example, \citet{Yokoyama2001} reported reconnection inflows in remote-sensing observations of a limb flare event, while recently \citet{Liu2013} and \citet{Takasao2012} reported clear observations of plasmoids in the trailing region of an erupting prominence, and in the region above the flare loops (see \fig{16}b). Other consequences of magnetic reconnection in coronal current layers after the onset of an eruption, such as turbulence \citep{Bemporad2008} and supra-arcades downflows \citep[see \eg][]{McKenzie2011,Savage2012,Savage2012b}, as well as hot high-speed plasma outflows \citep{WangT2007} have been reported over the years. Note that the motion of plasmoids (or plasma blobs in the corona) jetted away from the reconnection site (either upward or downard) is also reproduced in numerical simulations \citep[\eg][]{Barta2008}.

Radio observations also show drifting pulsating structures (DPS). These radio emissions differ from type II or type III bursts as they appear as intermittent bursts, and are believed to originate from the trapping of electrons inside plasmoids \citep[\eg][]{Karlicky2002}. Then, as these particles travel inside the structure, their radio emissions are directly related with the density and the size of the structure in which they are trapped. Indeed, analyses of DPS in radio emissions and remote-sensing images of plasmoids taken for example by AIA show a good correspondence in their occurences and locations \citep{Nishizuka2015}. In this paper, \citet{Nishizuka2015} also found a distribution of plasmoid width centered around $6.10^8$cm. According to \citet{Loureiro2012}, the maximum plasmoid size found in numerical simulations scaled as a tenth of the system size. Then, with plasmoids of the order of a megameter, one would expect a current layer of 0.1Mm. This seems to fit the observations of \citet{Takasao2012} (see \fig{16}b and their paper for an observation of what would be the coronal current layer structure). {We note that the correspondence between the plasmoids found in the corona, as indicated in the discussions above, and the plasmoids found in simulations, are yet to be confirmed. Indications may be given by the link between coronal plasmoids and small structures found in the interplanetary medium, as follows.}

Finally, as flux ropes are ejected in the interplanetary medium, it is possible to directly probe them with in situ instruments (such as with STEREO A/B, Wind or ACE at 1~AU). Such data have been for example used to obtain diagnostics of the structures of coronal mass ejections \citep[\eg][and references therein]{Zurbuchen2006}. A population of small flux ropes have been found to have a similar magnetic structure as interplanetary CMEs (twisted magnetic field), while following a completely different size distribution \citep[see][and \fig{16}d]{Janvier2014b}. Interestingly, the power-law of the distribution of small flux ropes is similar to the one found in numerical simulations of plasmoids (\fig{16}c). This indicates that similar processes in the formation of those plasma structures may occur in the current layers of the Sun's atmosphere. This is an interesting finding that supports the universality of the reconnection phenomenon and its consequences, from plasma experiments to numerical simulations and astrophysical plasmas.\\

%%%%%%%%%%%%%%%%%%%%%%%%%%%%%%%%%%%%%%%%%%%%%%%%%%%%%%%%%%%%%%%%%%%%%%%%%%%%%%%%%%%%%
\section{Discussions and Conclusion} %%%%%%%%%%%%%%%%%%%%%%%%%%%%%
\label{sect_Conclusion}

% {\S}{\bf --- Summary of the aim  } \\

%Anyone who was worked in understanding how magnetic reconnection must have gone through the same states I went through writing the present review. \red{Won't be in the text but... desperation is the word!} \\

Magnetic reconnection is a very complex phenomenon, that can be studied from various angles: analytical, numerical, observational, to study the change in topology, the energy conversion, MHD processes, bi-fluid or (gyro-)kinetic ones, etc. The different approaches are so numerous it is of course impossible to report all the findings in every branch of reconnection studies.

In the present review, the interest is focused on presenting this variety, as a guideline so as to remember that magnetic reconnection is difficult to understand in its ensemble, yet its studies and the advances along the years have provided powerful insights in the physics of solar flares. Especially, the advances provided by an increasing power in numerical calculations, as well as high spatial and temporal observations, have made possible the understanding of the magnetic reconfiguration that ensues from the reconnection mechanism. 

As such, we have reported in \sect{topology} how to understand magnetic reconnection from a topological point of view, looking at places where {the field line connectivity is discontinuous (such as null points and separators}, \sect{NP}) or where the magnetic connectivities are strongly distorted (\sect{QSLs}), leading under stress to the formation of current layers (\sect{currents}) where the magnetic energy is dissipated. For researchers interested in the mechanisms of the dissipation process, multiple ``layers'' can be considered: at MHD-scales, the evolution of boundary motions or spontaneous instability lead to the formation of magnetic islands (in a 2D cut), while at smaller scales, the reconnection rate is influenced by the behaviors of ions and electrons (\sect{dissipationprocess}). At scales that are of interest to understand the process of eruptive flares, numerical studies give us some insights on the structure and the evolution of the 3D current volume (\sect{currentlayer}), which can both be directly compared with observations (\sect{obscurrents}). As such, the aim of the paper is to provide a thorough understanding of the evolution of the standard model for flares in 3D.

The consequences of reconnection in 3D are multiple and have been reported in \sect{consequences}. We can first point out that the intrinsic 3D nature of magnetic reconnection leads to phenomena such as slipping motion of field lines that have recently been confirmed by high temporal cadence observations of the Sun's atmosphere (\sect{slipping}). The magnetic field 3D restructuration leads to typically observed evolutions of flare loops and flux ropes. For example, the strong-to-weak shear transition of flare loops as well as the growing envelope of flux ropes (\sect{flareloops}) are seen in both numerical studies and observations. Finally, observations at high spatial resolutions of the corona have revealed fine structures such as plasmoids. They have been the subject of many studies and provide comparisons between observations and numerical simulations of teared current layers  (\sect{plasmoids}). There are however other aspects that have not been touched upon in the present review, for reasons specified below.\\

%  {\S\bf Energy budgets}\\

The consequences of magnetic reconnection are related with the restructuration of the magnetic field as well as with the dissipation of magnetic energy. Then, one question that derives from the mechanisms of solar flares is what proportion of free magnetic energy (i.e. the difference between non-potential magnetic field and potential magnetic field energies) is actually converted in other sources of energy, and how this partition occurs.
%A direct observable consequence of magnetic reconnection is the plasma heating in the nearby region of magnetic field dissipation. This heat bulk appears in different forms. For example, a heat source in Hard X-rays  can be observed above the loop top region \citep{Masuda1994}, while coronal loops appear during the flare impulsive phase up to the end of the flare itself. These loops emit in SXR as well as in EUV light. The footpoints of the flare loops are also regions where HXR are found (Krucker etc).
\citet{Syrovatskii1966} described that the electric field is directed along $\vec{J}$, and thus will perform positive work on the charged particles, increasing their energy. It is this process that will convert the magnetic energy into the kinetic energy of the particles (dynamic dissipation). Such a mechanism is distinct from Joule dissipation as there is not a simple proportionality between the current density and the electric field. In \citet{VanHoven1973}, the authors discussed the energy release during the tearing instability, and showed that 10\% of the background energy can be released under the form of nonthermal energy.
More recently, \citet{Yamada2016} showed that in both plasma experiments and 2D simulations, about half of the magnetic energy is converted into particle energy, of which 2/3 is ultimately transferred to ions and 1/3 to electrons. 
Since those dedicated experiments and simulations are mainly focused on the dissipation layer, it would be interesting to extend such studies to their consequences in the large-scale changes seen during flares. For example, is there a link between the energy conversion seen at the dissipation scale, the energy carried by the energy spectrum of particles during flares/waves \citep[for example, high-frequency Alfv\'en waves energy conversion in the presence of turbulent magnetic fields, as investigated by][]{Lazarian1999} and the kinetic energy of the eruptive magnetic flux rope?

%Effect of magnetic islands on the electron dynamics: Drake Nature 2006 Oka 2010 ApJ (+ extension to flux ropes with LeRoux 2015 ApJ  Chen Nature 2007
%Electron acceleration: Drake 2005 PRL, Wan 2008 POP
%Electron Heating: Shay 2014

Bolometric energy (or total radiant energy) is a good proxy to infer the amount of energy that is released during flares under the form of direct heating and particle energy (emitting in, \eg, Hard X-rays). However, obtaining its values for flares is very difficult \citep{Kretzschmar2011}. From different flare events, \citet{Emslie2005} reported that (at least) 50\% of the magnetic energy was converted into the bolometric flare energy while the other half in the kinetic energy of the CME.
However, the estimation of the kinetic energy comes with a lot of hypotheses, such as the mass that is transported in CMEs. Such masses are generally inferred from remote-sensing coronal observations (such as with the LASCO instrument suites aboard SoHO), but can be mixed with the dense region created by a snow-plow effect of the erupting CME which is generally defined as the sheath region, well observed with in-situ instruments. This was for example reported by \citet{Feng2015} where the authors found that the effect of the solar wind ``snow-plow'' phenomenon was not negligible.

In numerical experiments, large-scale (either 2.5D or 3D) MHD simulations of eruptive flares such as in \citet{Amari2003b,Lynch2008,Reeves2010,Aulanier2012} show that the kinetic energy of the CME is roughly around 5\% and not more than 10\%, well below the amount proposed from observations. 
These discrepancies between numerical simulations and observations, as discussed by \citet{Aulanier2013}, must be resolved in the future both by improving the numerical scheme (dissipation mechanisms and atmospheric models in simulations) and the hypotheses/data constraints (the actual plasma mass transported in CMEs).\\

%Reconnection rates have been calculated in numerical simulations. In observations, determining the location of flare ribbons was shown to link with the parallel electric field \citep[integrated along magnetic field lines[]{Hesse2005}.  In particular, the reconnection rate and changes are directly related to its maximum value. Therefore, analysing the evolution of flare ribbons is a good tool to investigate the amount of flux that has been reconnected. Several authors have investigated as well as the reconnection rate \citep[see \eg][]{Xie2009,  Qiu2002, Isobe2005, Saba2006, Miklenic2007}. Magnetic reconnection rates deduced with such a technique were also found to correlate well with the acceleration phase of erupting flux ropes, as found by \citet{Jing2005}. However, one must be careful when considering such a simplified estimation of the reconnection rate, since we still need to investigate how the 3D nature of reconnection affects the reconnection rate. 

%  {\S\bf A plea for MHD simulations - to counter argue with people like H. Hudson}\\

Comments have been made that MHD simulations do not treat particles needed to explain the conversion of magnetic energy at ion and electron scales, as well as other nonlinear effects such as wave-particle interactions. However, as was shown throughout the text, the MHD paradigm is very good at explaining a varied range of observations. For example, we have seen that the strong gradients of the field-line mapping are related to locations of strong current densities. These correspond very well with regions of heating and particle acceleration such as null points and QSLs. Some may argue that the use of numerical techniques (as well as analytical ones) constrain the analysed configurations: those are over-simplified, when compared with the complexity of active regions and flare configurations. Nonetheless, studying such simplified models is still valuable, because it provides a quantitative test of the laws of physics we use to explain the underlying mechanisms of flares.

Furthermore, studying the large-scale restructuration of the magnetic field at MHD scales 
let us understand how flux ropes are created and are ejected in the heliosphere, Then, studies focused on their helicity content and $B_z$ orientation (important to estimate the impact on magnetospheres) are made possible by understanding the whole sequence of MHD evolution.
As pointed out by \citet{Forbes2000}, there is an enormous mathematical difficulty in solving the equations governing the motion of plasma and the behavior of magnetic fields. One the one hand, fluid equations have a high degree of nonlinearity, well illustrated by turbulent behaviors, and on the other hand, numerical methods cannot handle well (yet) the large gap between scales seen in solar flares, from the electron gyro-radius at $10^{-2}$m to the megameter range ($10^9$m) observed in CMEs. 
A solution would be to bridge different simulations, for example by providing MHD solutions as inputs in energetic particles models describing the motion of the particles in electromagnetic fields.\\

%      {\S\bf Discussion - limitation of MHD models and further developments} \\

In the GEM challenge (Geospace Environmental Modeling Magnetic Reconnection Challenge, see \sect{dissipationprocess}), the authors \citep{Birn2001a} found that different codes that include small scale physics are different from the MHD code. In particular, all the models that include the Hall effect in the generalised Ohm's law behave similarly (in terms of reconnection rate), and with a higher reconnection rate than the MHD model. It looks like the effect of the Whistler waves, brought in the dynamics by the Hall term, is important in modelling magnetic reconnection.
How this will affect the evolution of the magnetic field during solar flares, in 3D, is something that is still to be analysed. {Moreover, recent studies such as that of \citet{Bessho2005} show even when the Hall term is not present (canceled out, in the case of an electron-positron plasma), while fast reconnection still takes place, mediated by the pressure tensor (which off-diagonal terms produce an effect similar to spatially localised resistivity)}. Effects at electron scales therefore may provide invaluable insights in understanding the mechanisms solar flares. Finally, in full 3D systems, current layers are unstable to a wider range of instabilities than in 2D system, so this is something that remains to be looked at more deeply. 

Another great difficulty in understanding the mechanisms of flares reside in its timescale: during the impulsive phase of large flares, which lasts a few minutes, about $10^{25}$J of energy is released within a short amount of time. How can such amount of energy be converted into thermal and kinetic energy within the dissipation layer, which can be assumed to be of the order of $10^{15}$ km$^3$?.

Citing \citet{Gonzalez2016}, the editors wisely wrote ``One can also see that no consensus about [fundamental issues in magnetic reconnection] is presently available, thus indicating that the subject of magnetic reconnection remains open for further theoretical, computational, and observational studies.'' Nonetheless, reconnection during eruptive flares is an exciting field where progresses have allowed us to understand, but also predict  its underlying mechanisms. Dedicated missions such as the future Solar Orbiter will allow to better constrain the physical inputs of reconnection consequences, such as flux rope ejections, by directly probing the near environment of the corona.\\

\begin{acknowledgements}
The author would like to thank the two reviewers for their comments which help improve the manuscript. She would like to acknowledge the discussions she had at the International Scientific Seminar ``Structure and dynamics of the solar atmosphere: recent advances and future challenges'', Kavli Royal Society International Center, which resulted in the addition of some materials for the present manuscript. The author would also like to thank T. Passot, editor of JPP, for his patience in receiving the manuscript, P. D\'emoulin for his comments helping the revision of the text and C. Damiani for her support.
% We thank ... for reading carefully, and so, improving the manuscript.
%% Grants
\end{acknowledgements}

%\appendix
%\section{}\label{appA}
%This appendix contains sample equations in the JPP style. Please refer to the {\LaTeX} source file for examples of how to display such equations in your manuscript.

\bibliographystyle{jpp}
% Note the spaces between the initials
\bibliography{JPP-finalsubmission}

\end{document}